\documentclass[%
 aip,
  jcp
 amsmath,amssymb,
preprint,%
]{revtex4}

\usepackage[version=3]{mhchem} 
\usepackage{epsf}
\usepackage{ifthen}
\usepackage[usenames]{color}
\usepackage{amssymb,amsmath,amsbsy}
\usepackage{amsfonts}

\usepackage{lineno}
\usepackage{dsfont}
\usepackage{amsmath}
\usepackage{subfigure, epsfig}
\usepackage{graphics, graphicx}
\usepackage{bm}
\usepackage{natbib}

\begin{document}
\title[OC]
  {Quantum Optimal Control Theory for Solvated Systems}

\author{Marta Rosa}
\affiliation{Dipartimento di Scienze Chimiche, Universit\`a degli Studi di Padova, Padova, Italy}
\author{Gabriel Gil}
\affiliation{Dipartimento di Scienze Chimiche, Universit\`a degli Studi di Padova, Padova, Italy}
\affiliation{{\color{black}Instituto de Cibern\'etica, Matem\'atica y F\'isica, La Habana, Cuba}}
\author{Stefano Corni}
\email[Corresponding author: ]{stefano.corni@unipd.it}
\affiliation{Dipartimento di Scienze Chimiche, Universit\`a degli Studi di Padova, Padova, Italy}
\affiliation{Center S3, CNR Institute of Nanoscience, Modena, Italy}
\author{Roberto Cammi}
\email{roberto.cammi@unipr.it}
\affiliation
{Dipartimento di Scienze Chimiche, della Vita e della Sostenibilit\`a Ambientale, Universit\`a di Parma, Parma, Italy}





\begin{abstract}
 In this work, we extend the quantum optimal control theory of molecules subject to ultrashort laser pulses to the case of solvated systems, explicitly including the solvent dielectric properties in the system Hamiltonian. A reliable description of the solvent polarization is accounted for within the Polarizable Continuum Model (PCM). The electronic dynamics for the molecule in solution is coupled with the dynamics of the surrounding polarizable environment, that affects the features of the optimized light pulse. Examples on test molecules are presented and discussed to illustrate such effects. 
\end{abstract}

\maketitle
\section{Introduction}
Our capability to control the ``micro-world of chemistry and physics"\cite{Rabitz2002} lies on the possibility to selectively guide the dynamics of a quantum system through the interaction of the system with {\color{black} a suitable perturbing agent} (e.g., {\color{black} an external} laser pulse), in such a way that starting from a given initial state, the system is driven to a desired final state (e.g., a specific excited state, a specific product in a chemical reaction, {\color{black} a desired many-body quantum state}, etc). \cite{Tannor1985,Peirce1988,kosloff1989wavepacket,keefer2018pathways,deffner2014optimal,allen2017optimal,castro2019optimal}

Many efforts have been devoted to the control of nuclear dynamics, allowing to successfully achieve specific products ratios of chemical reactions and crystallization processes, photodissociation of molecules in gas phase, energy flow optimization in light-harvesting complex, 
and generally proving the method valuable in the conception of effective experiments \cite{henriksen2002laser,ma2002optimal,kehlet2004improving,kehlet2007optimal}. From the experimental point of view, studies mainly employed learning algorithms, where the external field is iteratively modified following an experimental feedback\cite{rabitz2000optimal}. 
Recently, femtosecond experiments on systems with complex multidimensional potential energy surfaces have been used together with theoretical ab initio quantum calculations to identify specific Hamiltonian information and optimal laser pulses, as well as to bring the system in the desired target state\cite{geremia2002optimal, daniel2003deciphering, accanto2017rapid}. \\
{\color{black} Application of optimal control theory in single molecule spectroscopy studies are of great interest, with the aim to identify ultrafast laser pulses to gain access to the intrinsic ultrafast molecular processes such as vibrational motions, excitation energy transfer, charge transfer, etc \cite{brif2010control,nuernberger2007femtosecond}. In all these possible molecular application, the target of the optimal control problem is a specific excited state of the molecule. In our study, we will indeed focus on problems where the target state is an electronic excited state, with the molecule at equilibrium bond length.}
From the theoretical side, the control of a quantum system relies on the quantum optimal control theory (QOCT). In this approach, the {\it optimal} laser pulse is
calculated a priori from the assumed system’s Hamiltonian\cite{Tannor1985,Peirce1988,shi1988optimal, kosloff1989wavepacket, zhu1998rapidly,werschnik2007quantum}.
Optimal control theory is widely applied to the study of molecular electronic states with ultrashort laser pulses \cite{klamroth2006optimal, werschnik2007quantum, krause2008dipole}, which require a short time scale (fs) to avoid nuclear relaxation and energy level rearrangements; with the application of optimal control theory it is possible to compute perfectly suited laser fields able to drive the system to the desired target state. Once the final shape of the pulse is obtained, the analysis of its polarization and frequencies can allow to understand the molecular behaviour and excitation mechanism\cite{klamroth2006optimal, krause2008dipole}. \\
The quantum optimal control theory has been successfully applied to gas phase molecules but, contrarily to the experimental approaches, its application to molecules in the presence of an external environment (e.g. a solvent) is more limited. Many studies focused on targeting specific vibronic molecular states in condensed phases using density matrix theory to introduce dissipative coupling with external bath \cite{ohtsuki2003quantum,beyvers2006optimal,keefer2018pathways}. 
In addition, concerning specifically the effect of solvation in optimal control problems, an effort was done by Keefe et al. to account for the molecular geometry fluctuations due to the electrostatic interaction with the molecular dipoles of the solvent, as certain structures may be stabilized or destabilized by the solvent presence\cite{keefer2018pathways,keefer2015multi}. In their study, they select different molecular geometries from a simulation performed with explicit solvent and calculate the vibronic excited state for all of them. The purpose of their multi target optimal control theory is then to find the optimal laser pulse which performs better on average for all the different structure, as all of them are present in the solvated system.\\ 

All these studies partially include the effect of the solvent on the control problem, and none is able to account at the same time  both the interaction of the molecular system  with the external control field and the polarization interaction of the same  system with the external medium (i.e. the so called reaction field problem). Furthermore, in the presence of an external medium surrounding the target molecules we have to face two additional problems regarding a coherent extension of the QOCT.  The first one is  due to the fact that in a condensed medium the electric field accessible to  the observer for  the control of the molecular system  is different with respect to the field acting locally on the system itself (i.e. the so called local field problem);\cite{Onsager1936} the second problem regards the nature of the excited electronic state (i.e the target states) of a solvated molecule  with respect to the case of an isolated molecules.\cite{tomasi2005quantum} In fact due to the interaction with the polarizable solvent the excited states are no more elements of the eigenfunction spectrum of a single electronic Hamiltonian but are eigenfunctions of different state-specific Hamiltonians. We remark that these characteristics  of the excited electronic state of molecules in a polarizable  environment reflects the non linear-nature of the molecular Hamiltonian associated to the quantum mechanical problem for these systems.  This is a characteristic already met with in quantum optimal control problems, although in rather different contexts (e. g. Bose-Einstein condensates). \cite{sklarz2002loading,hohenester2007optimal,mundt2009optimal} 

In this manuscript we present a generalization of the QOCT for solvated molecules in which all these effects are for the first time explicitly considered. The molecules in solution will be described within the Polarizable Continuum Model (PCM) for solvation.\cite{Miertus1981,Cammi1995,Cances1997,
cammi1998calculation,tomasi2002molecular,
tomasi2005quantum,cammi2005electronic}. The PCM is ``de facto" a standard  of contemporary computational quantum chemistry for the study of the molecular properties and processes in solution. PCM is an implicit solvation model in which the solvent is represented as a dielectric medium and the solute interacting with the medium and with external field of various nature and complexity can be described at the various standard theoretical level of quantum chemistry. 

The paper is organized as follows: in the ``Theory" part, we present the theoretical formalism; in particular in section \ref{sec:qoct} we review a suitable form of the QOCT for isolated molecules; in section \ref{sec:pcm_rev} we review the basic aspects of the PCM solvation model necessary for the  extension of the QOCT to molecules in solution; this extension is presented in section \ref{sec:PCM}.In sect. \ref{sec:algorithm} we discuss the algorithm implementing the QOCT approach.

In the ``Computational applications" part, we present the computational protocol and the numerical results of the QOCT for two molecules in solution, N-methyl-6-quinolone (MQ) and LiCN. In this part, we explore how the effect of the solvent modifies the optimal control process itself, in terms of the computational effort needed to solve the optimal control problem and the shape and amplitude of the final control pulse, and how the presence of the solvent modifies the final result, in terms of the state actually reached by the molecular system.

\section{Theory}
\subsection{Quantum Optimal Control Theory}\label{sec:qoct}
Quantum Optical Control Theory (QOCT) has been actively developed since the mid '80s.\cite{Tannor1985,Peirce1988,kosloff1989wavepacket} We review here one of the possible approaches to QOCT; following closely the works of Rabitz et al. \cite{shi1988optimal, zhu1998rapidly,werschnik2007quantum} we want to manipulate 
a suitably shaped laser pulse, in order to drive the system from an initial state at $t = 0$ where $\psi(0) = \psi_0$ to a final state $\psi(T) = \psi_T$ corresponding to a chosen final time $t = T$. The final state should maximize the expectation value of a chosen operator $\hat O$ acting on the system. In this work, $\hat O$ will be the projection operator on the desired excited state $\Psi$, i.e., $\hat{O}=|\Psi\rangle\langle\Psi|$.

\begin{equation}\label{eq:ex_value}
max_{\boldsymbol {\varepsilon}(t)}\;  O(T) = max_{\boldsymbol{ \varepsilon}(t)}\:  \langle \psi(T)| \hat O | \psi(T) \rangle 
\end{equation}
The molecule is treated quantum-mechanically, while $\boldsymbol {\varepsilon}(t)$ is the electric field associated to the laser pulse, that is the quantity an experimenter can directly control. 
The time dependent Schr\"odinger equation for the system is:
\begin{equation}\label{eq:sh}
{\it i} \frac{\partial\psi(t)}{\partial t} = [ \hat H_0 -  \boldsymbol {\varepsilon}(t)\boldsymbol{\hat{{\mu}}}]\psi(t)
\end{equation}

where $\hat H_0$ is the system Hamiltonian without the interaction with the external field, $\boldsymbol{\varepsilon}(t)$ is the external control field and $\boldsymbol{\hat{{{\mu}}}}$ is the dipole operator of the molecule. Eq. \ref{eq:sh} is explicitly written in the length gauge (other gauges are possible as well).\cite{Pipolo2014} Note that atomic units are used throughout this work.\\
The optimal field can be obtained maximizing the following constrained objective functional:
\begin{equation}\label{eq:J}\begin{split}
J = &\langle\psi(T)|\hat O| \psi(T)\rangle - \int_{0}^{T} \alpha(t)|\boldsymbol{\varepsilon}(t)|^{2} dt \\
&- \Bigg [ \int_{0}^{T} \langle \chi(t) | \left[ \frac{\partial}{\partial t} + {\it i} ( \hat H_0 - \boldsymbol{\varepsilon}(t)\boldsymbol{\hat{{\mu}}} )\right] | \psi(t) \rangle dt \Bigg ] \\
&- \Bigg [ \int_{0}^{T} \langle \left[ \frac{\partial}{\partial t} + {\it i} ( \hat H_0 - \boldsymbol{\varepsilon}(t)\boldsymbol{\hat{{\mu}})\right]} \psi(t) | \chi(t) \rangle dt \Bigg ] 
\end{split}
\end{equation}

where $\chi(t)$ is  
the Lagrange multiplier imposing that $\psi(t)$ satisfy the time dependent Schr{\"o}dinger equation at any time {\it t}, and $\int_{0}^{T}\alpha(t)|\boldsymbol{\varepsilon}(t)|^{2} dt$ is a positive function which plays the role of a penalty factor: the higher the laser fluence, the more negative its contribution to $J$. The time dependence of $\alpha(t)$ allows to enforce a given envelope to the laser pulse, e.g. penalizing too strong values at the beginning or end of the pulse.\\ 

To compute the stationary points of {\it J}, i.e. $\partial J=0$, we have to differentiate with respect to $\psi(t)$, $\chi(t)$ and $\boldsymbol{\varepsilon}(t)$. {\color{black} Details are given in Appendix A.}
From the condition $\partial_{|\psi\rangle}J$ = 0, $\partial_{|\chi\rangle}J$ = 0, $\partial_{\varepsilon}J$ = 0 come three coupled equations for the wave function $\psi(t)$, the Lagrange multiplier $\chi(t)$ and the field \cite{werschnik2007quantum}:

\begin{equation}\label{eq:wf}
{\it} \frac{\partial\psi(t)}{\partial t} = -{\it i}[\hat H_{0} - \boldsymbol{\varepsilon}(t)\boldsymbol{\hat\mu}]  \;\psi(t) \qquad \psi(0)=\psi_0
\end{equation}

\begin{equation}\label{eq:chi}
{\it} \frac{\partial\chi(t)}{\partial t} = -{\it i}[\hat H_{0} - \boldsymbol{\varepsilon}(t)\boldsymbol{\hat\mu}] \;\chi(t) \qquad \chi(T)=\hat O \psi(T)
\end{equation}

\begin{equation}\label{eq:field}
\boldsymbol{\varepsilon}(t) = -\frac{1}{\alpha(t)}{\it Im}[\langle\chi(t)|\boldsymbol{\hat\mu}|\psi(t)\rangle]
\end{equation}

Several algorithm have been implemented to numerically solve this problem. Here we adopt an iterative algorithm introduced by Rabitz and coworkers in ref. \cite{zhu1998rapidly}. When using this algorithm in vacuo, {\it J} has the property to increase monotonically. \\
Practically the algorithm builds upon a forward propagation of $\psi(t)$ starting with $\psi(0) = 0$ and a backwards propagation of $\chi(t)$ starting with $\chi(T) = \hat O \psi(T)$. The latter procedure has been interpreted in a rather transparent way by Tannor in his textbook.\cite{tannor2007introduction} The projector $\hat O$ eliminates from $\psi(T)$ the unwanted contributions from states other than $\Psi$, and the backward propagation is needed to find out what were the components of the field that gave rise to the desired portion of $\psi(T)$. 
More details on our practical implementation are given in Section \ref{sec:algorithm}.

\subsection{Polarizable Continuum Model}\label{sec:pcm_rev}
In this section we shall briefly review the Polarizable Continuum Model (PCM). PCM provides a suitable framework to perform quantum-mechanical calculations of molecules embedded in a dielectric environment including polarization effects\cite{tomasi2005quantum,cammi1998calculation,cammi2000calculation,tomasi2002molecular,cammi2005electronic}. PCM features comprise: (1) considering the environment as a continuum and infinite dielectric medium characterized by a frequency-dependent dielectric function $\epsilon(\omega)$; (2) tackling the molecule through quantum mechanics; (3) assuming the molecule inside a vacuum cavity --shaped according to the molecular geometry-- inserted within the otherwise homogeneous dielectric environment; and (4) the {\it reaction-} and {\it cavity-}field polarization of the medium are described in terms of apparent surface charges (ASCs) $\mathbf{q}$ placed on the cavity surface and depending on the molecular potential and the applied electric fields at the cavity surface, respectively. {\color{black} More specifically, the {\it reaction field} is the electric field due to the solvent polarization induced by the solute charge density, and will modify both the energy levels and the wavefunctions of the molecular states. In turn, this will affect the way the molecule interacts with light, e.g., by modifying the molecular transition dipole moments. Later on we shall see that the mutual solute-solvent polarization mediated by the reaction field has also direct effect in the laser driven solute state evolution. The {\it cavity field} (that will be discussed more in details below) is related to the modification of the electric field associated with the laser pulse inside the dielectric within the cavity hosting the molecule}

\begin{figure}[h!]
\centering
 \includegraphics[width=0.6\linewidth]{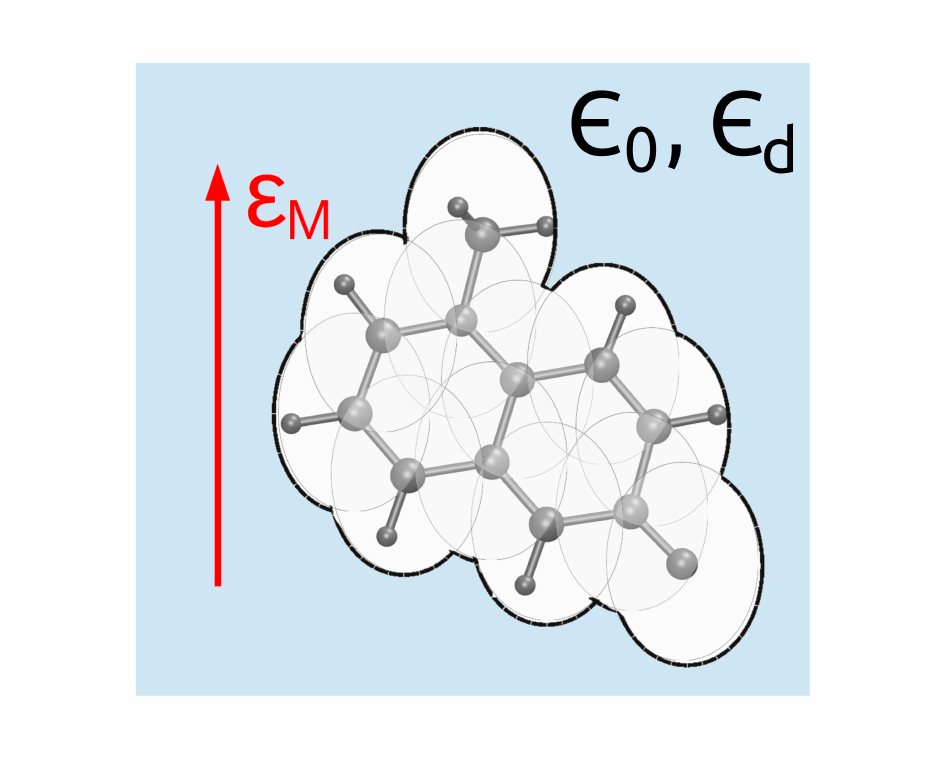}
\caption{{\color{black}Sketch of the PCM, with a molecular shaped cavity drawn for one of the molecule considered in this work.} $\boldsymbol{\varepsilon}_M$ is the Maxwell electric field in the dielectric, that is characterized by a static ($\epsilon_0$) and a dynamic ($\epsilon_d$) dielectric constant. {\color{black} At the boundary of the solute cavity, $\boldsymbol{\varepsilon}_M$ reaches the discontinuity of the dielectric function that steps from $\epsilon_0$ (or $\epsilon_d$) to 1. Polarization surface charges therefore appear, indicated as ${\mathbf q}_{cf}$ in the text, that modifies $\boldsymbol{\varepsilon}_M$ locally and in particular inside the solute cavity (cavity field effects).}}
\label{fig:pcm-cavity}
\end{figure}

The  time dependent Schr\"odinger equation in the PCM case can be obtained through a variational formulation that includes non-equilibrium effects \cite{Cammi1995b}, leading to a non-linear Schr\"odinger equation: \cite{sklarz2002loading}
\begin{equation}\label{eq:sh_pcm}
{\it i}\frac{\partial\psi(t)}{\partial t} = [ \hat H_0 + \boldsymbol{q}[\psi(t'<t)] \cdot \boldsymbol{\hat V} -  \boldsymbol {\varepsilon}_M(t)\boldsymbol{\hat{\bar\mu}}]\psi(t)
\end{equation}

where $\hat H^{0}$ is the Hamiltonian of the isolated molecule, $\boldsymbol{q}[\psi(t'<t)]$ represents the time dependent solvent polarization charges induced by the solute on the solvent (that in general depend on the entire history of the solute wavefunction $\psi(t')$ up to time $t$) placed on the boundary of the cavity, $\boldsymbol{\hat V}$ is the molecular electrostatic potential of the solute at the representative points on the cavity boundary;\cite{tomasi2005quantum,cammi1998calculation,pipolo2017equation} $\boldsymbol{\hat{\bar{\mu}}}$ is the effective electric dipole operator and the subscript M in the field ($\boldsymbol {\varepsilon}_M(t)$) reminds that this is the Maxwell field in the solvent. Effective dipole and Maxwell field will be discussed later on.

{\it Non equilibrium reaction field effects.} At the beginning of the simulated experiment (for $t\leq 0$), the solute is in its ground state and in equilibrium with the solvent. When the laser pulse has impinged the system (for $t>0$), the full system (solute and solvent) evolves according to the fast electric field oscillations of the laser. The slow degrees of freedom of the solvent cannot follow neither the laser pulse nor the solute density fluctuations in response to the laser field, and therefore, remain frozen in the initial configuration at equilibrium with the ground state of the solute. Only the fast degrees of freedom of the solvent are able to adapt to the perturbations, induced by the laser pulse propagation and the solute evolution. We incorporate this phenomenology by considering the following time dependent PCM equations for the polarization charges:\cite{Corni2014}

\begin{equation}\label{eq:q_pcm}
\boldsymbol{q}(t) = \int_{-\infty}^{\infty} \boldsymbol{Q}^{PCM} (t - t') \langle\psi(t') | \boldsymbol{\hat V} | \psi(t')\rangle dt'
\end{equation}
\noindent where ${\bf Q}^{PCM} (t - t')$ is the solvent response matrix, non-local in time and depending on the whole spectrum of the frequency-dependent dielectric permittivity of the medium.
In this work we perform the integral and rewrite the polarization charges  $\boldsymbol{q}(t)$ in Eq. (\ref{eq:q_pcm}) as a sum of two set of charges, following non-equilibrium treatments\cite{Cammi1995,Corni2014}: a dynamical charge term which follows the instantaneous polarization ($\boldsymbol{q}_d(t)=\boldsymbol{Q}^{PCM}_{d} \langle\psi(t) | \boldsymbol{\hat V} | \psi(t)\rangle$) and an inertial term  $\boldsymbol{q}_{in}$, constant in time, which is the response to the potential due to the ground state of the molecule ($\boldsymbol{q}_{in}=(\boldsymbol{Q}^{PCM}_0-\boldsymbol{Q}^{PCM}_{d}) \langle\Phi_0 | \boldsymbol{\hat V} | \Phi_0\rangle=\boldsymbol{q}_0(| \Phi_0\rangle)-\boldsymbol{q}_d(| \Phi_0\rangle)$):
\begin{eqnarray}\label{eq:q_pcm_split}
\nonumber \boldsymbol{q}(t) &=& \boldsymbol{Q}^{PCM}_{d} \langle\psi(t) | \boldsymbol{\hat V} | \psi(t)\rangle +
(\boldsymbol{Q}^{PCM}_0 - \boldsymbol{Q}^{PCM}_{d}) \langle\Phi_0 | \boldsymbol{\hat V} | \Phi_0\rangle=\\
&=& \boldsymbol{q}_d(t) + \boldsymbol{q}_{in}=\boldsymbol{q}_d(t) + \boldsymbol{q}_0(| \Phi_0\rangle)-\boldsymbol{q}_d(| \Phi_0\rangle) \end{eqnarray}
$\boldsymbol{Q}^{PCM}_0$ and $\boldsymbol{Q}^{PCM}_d$ are the PCM response matrices obtained by the static $\epsilon_0$ and the dynamic $\epsilon_d$ dielectric constants, respectively. As done before\cite{pipolo2017equation}, we can also rewrite $\boldsymbol{q}(t)$ as:
\begin{eqnarray}\label{eq:q_pcm_split_2}
\boldsymbol{q}(t) &=& \boldsymbol{q}_0(| \Phi_0\rangle) +\Delta\boldsymbol{q}_d(t)\\
\Delta\boldsymbol{q}_d(t)&=&\boldsymbol{q}_d(t)-\boldsymbol{q}_d(| \Phi_0\rangle)
\end{eqnarray}

We also define charge operators $\boldsymbol{\hat{q}}_d=\boldsymbol{Q}^{PCM}_{d}\boldsymbol{\hat V}$ and $\Delta\boldsymbol{\hat{q}}_d=\boldsymbol{Q}^{PCM}_{d}\boldsymbol{\hat V}-\boldsymbol{q}_d(| \Phi_0\rangle)$ in such a way that:
\begin{equation}
\boldsymbol{q}_d(t)=\langle\psi(t) |\boldsymbol{\hat{q}}_d  | \psi(t)\rangle
\end{equation}
\begin{equation}
\Delta\boldsymbol{q}_d(t)=\langle\psi(t) |\Delta\boldsymbol{\hat{q}}_d  | \psi(t)\rangle
\end{equation}

{\it Cavity field effects.} In Eq.\ref{eq:sh_pcm}, it appears  the effective dipole $\boldsymbol{\hat{\bar{\mu}}}= \boldsymbol{\hat\mu} + \boldsymbol{\hat{\tilde{\mu}}}$ which includes the cavity field effect\cite{pipolo2014cavity, gil2019non}. The latter effect can be cast by
\begin{equation}
-\boldsymbol{\hat{\tilde{\mu}}}\cdot\boldsymbol{\varepsilon}(t) = \mathbf{q}_{\mathrm{cf}}[\boldsymbol{\varepsilon}_M(t)]\cdot\hat{\mathbf{V}} 
\end{equation}
\noindent  where $\mathbf{q}_{\mathrm{cf}}[\boldsymbol{\varepsilon}_M(t)]$ are the polarization charges due to the control electric field $\boldsymbol{\varepsilon}_M(t)$ {\color{black} that develop to satisfy the electrostatic boundary conditions at the dielectric function discontinuity between the solvent and the molecular cavity, see Fig.(\ref{fig:pcm-cavity}). $\boldsymbol{\varepsilon}_M(t)$} is a so-called Maxwell field, i.e., a field that could be measured in the dielectric in the absence of the solute cavity. $\boldsymbol{\varepsilon}_M(t)$ is therefore the field that can be directly controlled by shaping the incident laser pulse. The $\mathbf{q}_{\mathrm{cf}}[\boldsymbol{\varepsilon}_M(t)]$ are given by a PCM-like equation:\cite{cammi1998calculation,cammi2000calculation,gil2019non}
\begin{equation}\label{pcm_eq_lf}
\mathbf{q}_{\mathrm{cf}}[\boldsymbol{\varepsilon}_M(t)]=\boldsymbol{Q}_{cf}\boldsymbol{V}_{\varepsilon_M}(t)
\end{equation}
where $\boldsymbol{Q}_{cf}$ is the cavity-field analogue of $\boldsymbol{Q}^{PCM}_d$ and $\boldsymbol{V}_{\boldsymbol{\varepsilon_M}}(t)=-\boldsymbol{\varepsilon}_M(t)\cdot \boldsymbol{r}$ is the potential associated with the control Maxwell electric field in the long-wavelength limit. The field produced by the charges $\mathbf{q}_{\mathrm{cf}}$ generalizes to PCM the cavity field originally introduced by Onsager for spherical cavities.\cite{Onsager1936} {\color{black} We also remark that within the dielectric the relation between the intensity of the propagating electromagnetic radiation and the field $\boldsymbol{\varepsilon}_M(t)$ is slightly different than in vacuo, as for a given $\boldsymbol{\varepsilon}_M(t)$, the intensity in the dielectric is a factor $n$ ($n$ is the refractive index) smaller than what would be in vacuo for the same field.} \\

{\it Ground and excited state description in PCM.} The state vector $|\psi(t)\rangle$ can be approximated in terms of a many-electron basis set 
\begin{equation}\label{eq:wf}
|\psi(t)\rangle = \sum_{I} C_{I}|\Phi_I\rangle
\end{equation}
\noindent where $|\Phi_0\rangle$ is the ground state while the others correspond to excited states. The excited states are generated applying an excitation operator to the reference state $|\Phi_{0}\rangle$, promoting the electrons from the occupied orbitals to the vacant ones.
The accuracy of the Schr\"odinger equation in solution depends on the choice of the basis set in Eq. \ref{eq:wf} expansion. For computational simplicity and feasibility, in this work we use as the reference the state given by the Hartee-Fock single determinant of the molecular solute, under a regime of equilibrium solvation ($|\Phi_{0}\rangle$  = $|{HF}\rangle$). The excited states $|\Phi_{I}\rangle$ are given by a configuration interaction expansion limited to single excitations (CIS) \cite{pipolo2017equation}. In particular, the quantum mechanics calculation providing the description of the molecule (i.e., the energy and ordering of its electronic states) is performed using the solvent polarization equilibrated with the solute ground state, which means that we are describing the system immediately after the excitation, when the solvent did not have time to equilibrate with the new electronic configuration of the solute. This is called the frozen solvent approximation, and we identify the frozen solvent excited states as $|\Phi_I^{fro}\rangle$.

More in details, the Hartree-Fock wavefunction  $|{HF}\rangle $ is obtained from  the solution of the Hartree-Fock equations involving the following Fock matrix (in the molecular orbitals basis):
\begin{equation}
F_{pq}^{PCM}=F_{pq}^{0}+{ {\bf q}}({|{HF}\rangle })\cdot{\bf V}_{pq}
\label{eq:fock}
\end{equation}
$F_{pq}^{0}$ are the matrix elements of the Fock operator for the isolated system; and ${ {\mathbf q}}({|HF\rangle})$  is defined by:
\begin{equation}
{ {\bf q}}({|{HF}\rangle })= {\bf Q}_0^{PCM} \langle {HF}|\hat{\bf V}|{HF}\rangle 
\end{equation}

The compositions and energies of the CIS frozen excited states $|\Phi_I^{fro}\rangle $ are obtained by solving, in the space spanned by the Hartree-Fock determinant and by the  single excited determinants, the time-independent Schr\"{o}dinger equation for  the molecular solute in the presence of the fixed Hartee-Fock polarization charges:
\begin{equation}
\left [ \hat{H}^0+{  {\bf q}}(|{HF}\rangle)\hat{\bf V} \right ]|\Phi_I^{fro}\rangle =E_I|\Phi_I^{fro}\rangle~. 
\label{eq:ei}
\end{equation}

When the frozen approximation is relaxed, i.e., the solvent is let free to equilibrate with a given excited state. Each excited state is obtained as a stationary state of the functional:\cite{Cammi1995,cammi2005electronic}

\begin{equation}
\mathcal{G}[\Phi_I^{eq}]=\langle \Phi_I^{eq}|\left [ \hat{H}^0+ {\bf q}(\Phi_I^{eq})\hat{\bf V} \right] |\Phi_I^{eq}\rangle-\frac{1}{2} \langle \Phi_I^{eq}| {\bf q}(\Phi_I^{eq})\hat{\bf V} |\Phi_I^{eq}\rangle
\end{equation} 
 
 that yields a time-independent Schr\"odinger equation with a state-specific Hamiltonian:
 \begin{equation}
 \label{eq:h_state}
 \left[ \hat{H}^0+ {\bf q}(|\Phi_I^{eq}\rangle)\hat{\bf V} \right] |\Phi_I^{eq}\rangle = E^{eq}_I |\Phi_I^{eq}\rangle
 \end{equation}
 
  The excited state $|\Phi_I^{eq}\rangle$ will therefore be different from  $|\Phi_I^{fro}\rangle$. {\color{black} For example, } the frozen solvent states remain a good basis set for the new excited states, \cite{cammi2005electronic} i.e, in general it will be possible to write:
\begin{equation}\label{eq:target}
|\Phi_I^{eq}\rangle = C_0|HF\rangle + C_1|\Phi_1^{fro}\rangle + ... + C_I|\Phi_I^{fro}\rangle + ... + C_N|\Phi_N^{fro}\rangle
\end{equation} 
The expansion coefficients $C_0$, $C_1$, etc. are obtained by solving Eq.(\ref{eq:h_state})with a self consistent procedure:
\begin{enumerate}
    \item a specific frozen solvent excited state $|\Phi_I^{fro}\rangle$ is chosen as the first approximation to the solvated excited state $|\Phi_I^{eq}\rangle$;
    \item the polarization charges $\boldsymbol{q}=\boldsymbol{Q}_0^{PCM}\langle\Phi_I^{eq}|\boldsymbol{\hat V}|\Phi_I^{eq}\rangle$ ( $\boldsymbol{q}=\boldsymbol{Q}_0^{PCM}\langle\Phi_I^{fro}|\boldsymbol{\hat V}|\Phi_I^{fro}\rangle$ at the first step) corresponding to the approximated solvated excited state ($|\Phi_I^{fro}\rangle$ at the first step) are computed;
    \item the Hamiltonian of the solvated molecule (without the laser) is represented in the basis set of the frozen solvent excited states, i.e., $\langle \Phi_K^{fro}|\hat{H}|\Phi_K'^{fro}\rangle = \langle \Phi_K^{fro}|\hat{H}_0|\Phi_K'^{fro}\rangle+\boldsymbol {q}\langle\Phi_K^{fro}|\boldsymbol{\hat{V}}|\Phi_K'^{fro}\rangle$, and we diagonalize it;
    \item out of the set of states obtained from the diagonalization, the state with the largest overlap with the approximated solvated excited state is selected as the updated approximation for it;
    \item loop from 2) to 5) until the desired accuracy (e.g., in terms of solvated excited state energy) is achieved. 
\end{enumerate}
This computational strategy to find solvated excited states is similar in spirit to the standard approach employed for solvated ground states. The key change is that instead of assuming a Slater determinant ansatz for the solvated ground state wavefunction and optimizing a linear combination of single-particle states, here we optimize a linear combination of frozen solvent excited (multi-particle) states.

\subsection{Quantum Optimal Control Theory in the Polarizable Continuum Model}\label{sec:PCM}
In this section, we discussed how to modify the optimal control algorithm described previously for its application in the case of a solvated molecule, with a PCM description of the environment.
Including the PCM terms in the optimal control problem, the {\it J} functional then becomes:
\begin{equation}\label{eq:J_PCM}\begin{split}
J^{PCM}= & \langle\psi(t)|\hat O|\psi(t)\rangle - \int_{0}^{T} \alpha(t)|\boldsymbol{\varepsilon}_M(t)|^{2} dt\\
&- \Bigg [ \int_{0}^{T} \langle \chi(t) | \left[ \frac{\partial}{\partial t} + {\it i} [ \hat H^{0} + (\langle\psi(t)|{\bf \hat q}_d | \psi(t) \rangle+{\bf q}_{in}) \cdot {\bf \hat V} - \boldsymbol{\varepsilon}_M(t)\boldsymbol{\hat{\bar{\mu}}}] \right] | \psi(t) \rangle dt \Bigg ] \\
&- \Bigg [ \int_{0}^{T} \langle \left[ \frac{\partial}{\partial t} + {\it i} [ \hat H^{0} + (\langle\psi(t)|{\bf \hat q}_d | \psi(t) \rangle+{\bf q}_{in}) \cdot {\bf \hat V}- \boldsymbol{\varepsilon}_M(t)\boldsymbol{\hat{\bar{\mu}}}]\right] \psi(t) | \chi(t) \rangle dt \Bigg ] \end{split}
\end{equation}

We proceed similarly to what we did in vacuo, differentiating $J^{PCM}$ with respect $\psi(t)$, $\chi(t)$ and $\boldsymbol{\varepsilon}_M(t)$. {\color{black} Details are given in Appendix A.}
The three final coupled equations are:

\begin{equation}\label{eq:wf_pcm}
{\it} \frac{\partial\psi(t)}{\partial t} = -i[\hat H^{0} + (\langle \psi(t)|{\bf \hat q}_d | \psi(t) \rangle +{\bf q}_{in}) \cdot {\bf \hat V} - \boldsymbol{\varepsilon}_M(t)\boldsymbol{\hat{\bar{\mu}}}] \psi(t) \qquad \psi(0)=\psi_0
\end{equation}

\begin{equation}\label{eq:chi_pcm}\begin{split}
{\it} \frac{\partial\chi(t)}{\partial t} = &-i[\hat H^{0} + (\langle \psi(t)|{\bf \hat q}_d | \psi(t) \rangle +{\bf q}_{in}) \cdot {\bf \hat V} - \boldsymbol{\varepsilon}_M(t)\boldsymbol{\hat{\bar{\mu}}}] \; \chi(t)\\
&- i\langle \chi(t)|{\bf \hat V} | \psi(t) \rangle \cdot {\bf \hat q}_d \; \psi(t) + i\langle \psi(t)|{\bf \hat V} | \chi(t) \rangle \cdot {\bf \hat q}_d \; \psi(t)\qquad \chi(T)=\hat O \psi(T)
\end{split}
\end{equation}

\begin{equation}\label{eq:field_pcm}
\boldsymbol{\varepsilon}_M(t) = -\frac{1}{\alpha(t)}{\it Im}[\langle\chi(t)|\boldsymbol{\hat{\bar{\mu}}}|\psi(t)\rangle]
\end{equation}

The PCM forward propagation equation has an additional term with respect to the in vacuo case, straightforwardly related to the additional $\langle\psi(t)|{\bf \hat q}_d | \psi(t) \rangle \cdot {\bf \hat V}$ term in the Hamiltonian. The backward propagation, on the other hand, has a more complex dependence on both  $\boldsymbol{\hat q}_d$ and  $\boldsymbol{\hat V}$ and a direct dependence on $\psi(t)$, {\color{black} which is a consequence of the $\langle\psi(t)|{\bf \hat q} | \psi(t) \rangle$ term, which generates extra terms in the equations (see Appendix \ref{sec:a2}).} Finally, eqs.(\ref{eq:wf_pcm}-\ref{eq:chi_pcm}) can be rewritten in terms of the HF charges $\boldsymbol{q}(|HF\rangle)$ as:
\begin{equation}\label{eq:wf_pcm_2}
{\it} \frac{\partial\psi(t)}{\partial t} = -i[\hat H^{0} +{\bf q}(|HF\rangle) \cdot {\bf \hat V}+ \langle \psi(t)|\Delta{\bf \hat q}_d | \psi(t) \rangle  \cdot {\bf \hat V} - \boldsymbol{\varepsilon}_M(t)\boldsymbol{\hat{\bar{\mu}}}] \psi(t) \qquad \psi(0)=\psi_0
\end{equation}

\begin{equation}\label{eq:chi_pcm_2}\begin{split}
{\it} \frac{\partial\chi(t)}{\partial t} = &-i[\hat H^{0} +{\bf q}(|HF\rangle) \cdot {\bf \hat V}+ \langle \psi(t)|\Delta{\bf \hat q}_d | \psi(t) \rangle  \cdot {\bf \hat V} - \boldsymbol{\varepsilon}_M(t)\boldsymbol{\hat{\bar{\mu}}}] \; \chi(t)\\
&- i\langle \chi(t)|{\bf \hat V} | \psi(t) \rangle \cdot {\bf \hat q}_d \; \psi(t) + i\langle \psi(t)|{\bf \hat V} | \chi(t) \rangle \cdot {\bf \hat q}_d \; \psi(t)\qquad \chi(T)=\hat O \psi(T)
\end{split}
\end{equation}
which are the equations effectively implemented, as they are the most convenient when the wavefunction basis set defined by Eq.(\ref{eq:ei}) is used.

\subsection{Optimization algorithm in vacuo and in solution}\label{sec:algorithm}
\begin{figure}[h!]
\centering
 \includegraphics[width=0.8\linewidth]{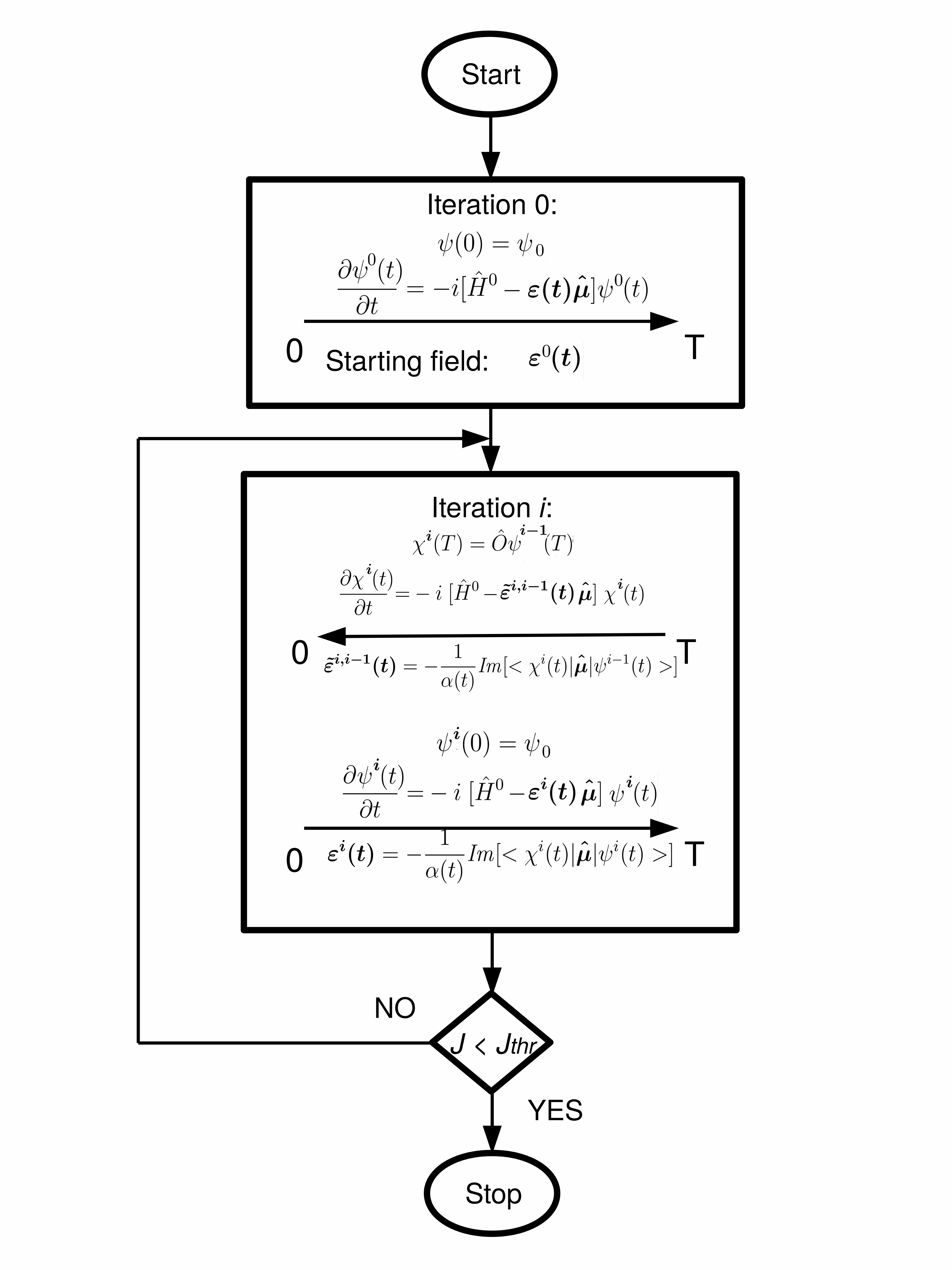}
\caption{Flow chart of the optimal control algorithm in vacuo. A similar procedure, with the PCM Hamiltonian, is performed for the system in implicit solvent.}
\label{sc:flow}
\end{figure}

Many algorithms were proposed to solve the optimal control problem.\cite{kosloff1989wavepacket,tannor2007introduction,keefer2018pathways} Here, we adopt the iterative algorithm introduced by Rabitz et al. \cite{zhu1998rapidly} briefly described in the following. In our code is also implemented a second algorithm described in Ref. \cite{ohtsuki2004generalized}, which rely on a very similar scheme. Each iteration of the algorithm but the first is made of a backward propagation of $\chi(t)$ and a forward propagation of $\psi(t)$, under the influence of two different fields (Scheme \ref{sc:flow}).\\
During the first iteration, which is usually identified as iteration 0, $\psi(t)$ is propagated from $t=0$ to $t=T$ under the effect of an initial guess for the electric field of the desired shape (e.g. {\it $\pi$-pulse}, constant,..). Values of $\psi(t)$ are stored at each time step. \\
Iteration 1 (and all the following iterations) starts with the evaluation of {\it J} or $ \langle\psi(T) |\hat O |\psi(T)\rangle$. If the desired level of accuracy (in terms of the value of $\Delta {\it J} = {\it J^1 - J^0} $, or the discrepancy between the target and the present value of $ \langle\psi(T) |\hat O |\psi(T)\rangle$)) is not achieved, $|\chi(T)\rangle$ is set equal to $\hat O |\psi(T)\rangle$ and a backward propagation is started. At each time step the {\it backward field} $\boldsymbol{\tilde{\varepsilon}_M}$ is calculated. This field is a {\it support} field, not the one we are aiming to tune with the optimal control algorithm, and its purpose is only to allow the backward propagation of $\chi(t)$. 
\begin{equation}\label{eq:bwd_field}
\boldsymbol{\tilde{\varepsilon}^{1,0}_M(t)}=-\frac{1}{\alpha(t)}{\it Im}[\langle\chi^{1}(t)|\boldsymbol{\hat{\bar{\mu}}}|\psi^{0}(t)\rangle]
\end{equation}
where the apex in Eq. \ref{eq:bwd_field} means that $\chi^{1}(t)$ is calculated at iteration 1 while $\psi^{0}(t)$ is calculated at iteration 0.\\
Once all the values of $\chi^{1}(t)$ are calculated and stored, one can propagate $\psi^{1}(t)$ forward, calculating $\boldsymbol{\varepsilon}_M(t)$ at each step as:
\begin{equation}\label{fwd_field}
\boldsymbol{\varepsilon}^1_M(t)=-\frac{1}{\alpha(t)}{\it Im}[\langle\chi^{1}(t)|\boldsymbol{\hat{\bar{\mu}}}|\psi^{1}(t)\rangle]
\end{equation}

where now $\chi^{1}(t)$ and $\psi^{1}(t)$ are both calculated at iteration 1.
The procedure is iterated until the desired accuracy on $\langle\psi(T)|\hat O |\psi(T)\rangle$ is obtained. 

This algorithm is immediately extendable to PCM, modifying the propagation equations with the additional PCM terms. In particular, as we have already mentioned, there is an additional term in the propagation equation of $\chi(t)$ that is due to an extra nonlinear dependence. Such term does not represent a problem for the algorithm in our particular case, since  backward propagation for $\chi(t)$ has access to the full time-dependence of $\psi(t)$  once performed the forward propagation. In general, nonlinearity has been already tackled with success in optimal control problems for Bose-Einstein condensate\cite{sklarz2002loading,hohenester2007optimal,mundt2009optimal}. There the non-linear term is local in space ($\propto |\psi(t)|^2$), while the PCM non-linear term is global, i.e., integrated in space. 
However, it should be noted that while the in vacuo algorithm was demonstrated to lead to a monotonic increase of {\it J}\cite{ohtsuki2004generalized}, the additional terms in the $\chi$ propagation equation do not allow the demonstration used in the in vacuo case. Practically, as it is shown in Section \ref{sec:results}, for the calculations we performed in solvent, which means at least for the specific molecules and parameters we consider, {\it J} behaves monotonically in solution as well (Sec. \ref{sec:results}).\\

\section{Computational applications}\label{sec:application}
In this section we present numerical applications of the QOCT-PCM method to the study of laser pulse for the optimal population of selected excited states of two molecular solvated systems: N-methyl-6-quinolone (MQ) and LiCN.\\

MQ is an interesting system with peculiar photo-physical properties,\cite{PerezLustres2005} already studied in the framework of optimal control in vacuo in Ref. \cite{klamroth2006optimal}. As stated there, the electrical pulse to excite this system to the first excited state must be very short (approx 6 fs.) to avoid nuclear relaxation. A {\it $\pi$-pulse} (i.e., a light pulse able to take all the ground state population in the excited state for a two level system)\cite{werschnik2007quantum} of such duration was shown to be ineffective to obtain the desired selectivity on the target excited state. Optimal control in vacuo already proved to solve this issue in Ref. \cite{klamroth2006optimal}, and we extend it to the solvent case. 

LiCN molecule has an electronic structure strongly affected by the presence of the solvent, and it is interesting to show how to deal with the choice of the target excited state in this particular case, and to study the differences in behaviour and performances of the optimal control algorithm. Moreover, LiCN was already used as a model system to test the extension of PCM to real time quantum approaches (real time time dependent functional theory, RT-TDDFT and time dependent CIS)\cite{Pipolo2014,Corni2014,pipolo2017equation} The present is a further extension of those works. We chose acetonitrile as solvent for both systems.

\subsection{Computational protocol}
The two molecular systems MQ and LiCN were treated at the same level of theory: the structures are obtained relaxing the geometry with Gaussian G09\cite{frisch2009gaussian} using a 6-31G(d) basis set at the Hartree-Fock level of theory. The molecular shaped cavity, see figure \ref{fig:pcm-cavity}, is made by the union of spheres centered on the heavy atoms of the molecules, with radii equal to  the  atomic  van  der  Waals  radii scaled by a factor of 1.2.

For LiCN the vdW radii are the ones reported  in  Ref. \cite{pipolo2017equation} [(Li = 2.17 \AA, C=2.04 \AA, and N = 1.83 \AA)] while for MQ they are obtained from a G09 ground state calculation in acetonitrile and are: $C_{aromatic}$ = 2.13 {\AA} and 1.93 {\AA}, $C_{CH_3}$ = 2.53 {\AA}, $N$ = 1.83 {\AA}, $O$ = 1.75 {\AA}. The parameters for the dielectric function of acetonitrile are $\epsilon_0$=35.84 and $\epsilon_d$=1.806 and are taken from Ref. \cite{pipolo2017equation}.\\
The CIS excited electronic states of MQ and LiCN in vacuo and in the presence of the PCM frozen solvent reaction field have been performed with a locally modified version of GAMESS package\cite{schmidt1993general}. The many-electron basis set $|\Phi_I\rangle$ is limited to the Hartree-Fock ground state and to the lowest 15 CIS excited states determined in vacuo and in the presence of the PCM frozen reaction field, see e.g., Fig. \ref{fig:quin-energy} for MQ. To check the suitability of using 15 excited states, we performed some test optimal control procedures both for MQ and LiCN molecules with 30 excited states, and the results obtained were equivalent both in terms of optimal field and final state of the molecule.

In our implementation the wavefunction $\psi(t)$ (and the Lagrange multiplier $\chi(t)$) are propagated through a first order Euler method, and $\psi(t)$ and $\chi(t)$ are not explicitly normalized in forward and backward propagation as the chosen time step (0.001 a.u., $\approx$ 2*$10^{-5}$ fs, unless differently stated) is small enough to assure an acceptable conservation of the wavefunction normalization. For $\chi(t)$, this allows to keep numerically consistent the information obtained from the condition $|\chi(T)\rangle=\hat O |\psi(T)\rangle$. {\color{black} The choice of Euler propagator was done to keep the procedure as simple as possible and focus on the extension to implicit solvent. We did not experience specific issues with this propagator; anyway more efficient propagators can be implemented and actually has been used for similar gas-phase algorithm (e.g., operator splitting technique in ref. \onlinecite{klamroth2006optimal}}).

It is also possible to choose between three different shapes for $\alpha(t)$:

\begin{equation}\label{eq:alpha}
\alpha(t) =
\begin{cases}
\alpha 0 \qquad  & =\alpha_{const}(t) \\
\frac{\alpha 0}{\sqrt(sin(\frac{\pi t }{T})} \qquad  & =\alpha_{sin}(t)\\
\frac{\alpha 0}{e^{{\left[\frac{-(t - t_0)}{\delta t_s}\right]}^{12}}} \qquad  & =\alpha_{smooth}(t) \\
\end{cases}
\end{equation}

where $T$ is the duration of the laser pulse (250 a.u., $\approx$ 6 fs, in all the simulations)\cite{klamroth2006optimal}. The $\alpha_{sin}(t)$ shape promotes sinusoidal envelope, with small values of the field amplitude at the beginning and at the and of the time interval, while  $\alpha_{smooth}$ shape is the one used in Ref. \cite{klamroth2006optimal}; this shape is a regularized version of a constant amplitude field which is suddenly turned on, slightly {\it smoothed} at the beginning and end of the pulse. In atomic units, $\alpha(t)$ and thus $\alpha_0$ are given in units of $e^2a_0^2\hbar^{-1}E_h^{-1}$.

In the present implementation of the QOCT-PCM optimization algorithm (see scheme in Fig. \ref{sc:flow}), the electric field in the first iteration $\boldsymbol{\varepsilon}^0$ has been selected on the basis of the following considerations. If the target state has well defined characteristics, one sensible choice would be to choose a $\pi$-pulse with an appropriate value of the frequency, able to guarantee a (partial) population inversion. Nevertheless, if the target state is a linear combination of excited states, as it can happen e. g. when targeting an excited state of the molecule in solution written as a combination of frozen solvent excited states, there is no point in choosing a preferred polarization for the starting field. On the contrary, depending on the cases, such a choice can slow down the optimal control algorithm. With the aim of obtaining general results and easing the comparison in vacuo vs. PCM, we performed all our calculations with constant starting fields oriented along the direction (111), with amplitude 0.01 a.u. ($\boldsymbol{\varepsilon}^0_{0.01}$=(0.01, 0.01, 0.01) a.u.), unless differently stated. 1 a.u. for the electric field is 1 $E_he^{-1}a_0^{-1}$.  The effect of different values of the amplitude of the starting
electric field $\boldsymbol{\varepsilon}^0$ has also been explored.

\subsection{Results and discussion}\label{sec:results}
\subsubsection{N-methyl-6-quinolone (MQ)}
\begin{figure}[h!]
\centering
 \includegraphics[width=0.5\linewidth]{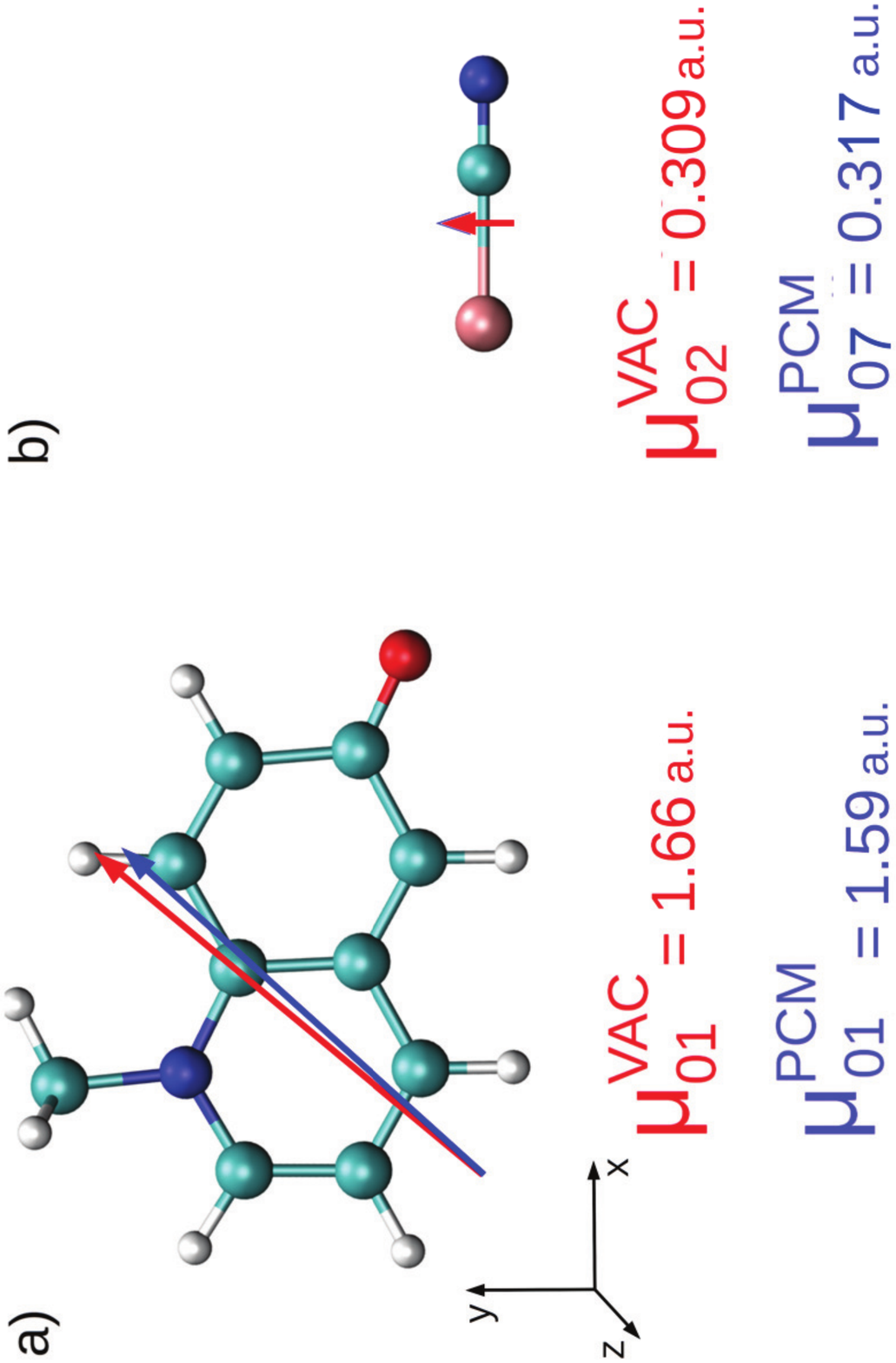}
\caption{N-methyl-6-quinolone (MQ) and LiCN molecules, superimposed with their transition dipoles for a) 0-1 transition, vacuo and PCM, and b) 0-2 transition in vacuo and 0-7 transition in PCM. In PCM the $|7^{fro}\rangle$ excited state is the main component of the $|7^{eq}\rangle$ state. Other dipole moments are not shown for clarity.}
\label{fig:molecules}
\end{figure}

\begin{figure}[h!]
\centering
 \includegraphics[width=0.5\linewidth]{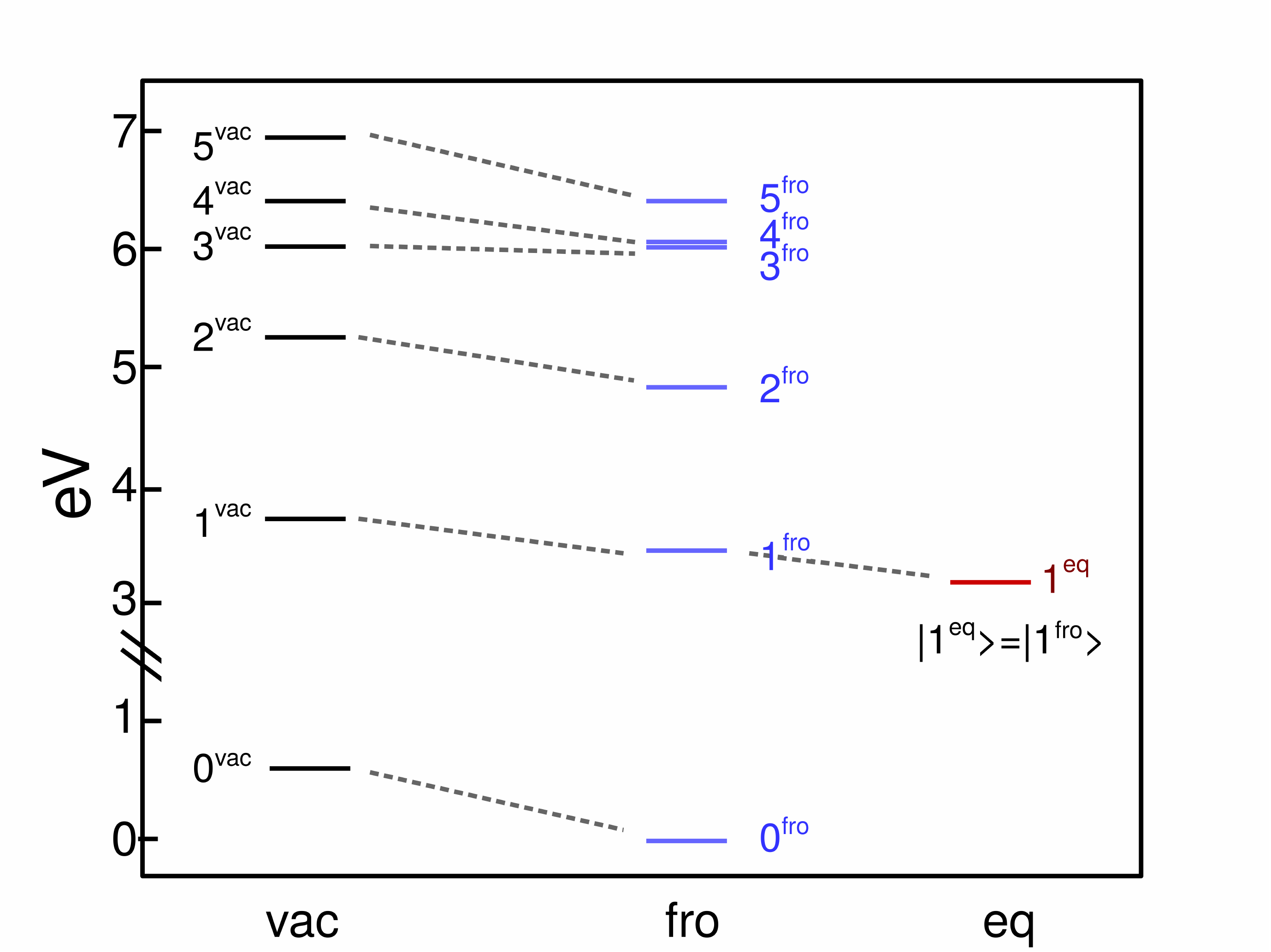}
\caption{Energies  (gas-phase)  and  free  energies  (in  solvent)  of  the  ground and the lowest excited states for MQ in vacuo (black) and within the frozen solvent approximation (blue) and free energy of the state chosen as target for the optimal control problem after the equilibration of the solvent with the excitation.  All the values refer to the equilibrium free energy of the ground state in solution which is therefore set to zero.}
\label{fig:quin-energy}
\end{figure}

We start performing a set of calculations with different parameters for the isolated molecule, to asses the behavior of the OC algorithm as implemented by us. Figure \ref{fig:molecules} a) shows QM molecule and the value and direction of the transition dipole for 0-1 transition, while  Figure \ref{fig:quin-energy} shows the energy levels of MQ molecule.\\

The optimal control problem has been solved, using different laser pulses in terms of shape and amplitude\cite{werschnik2007quantum}. The {\color{black} parameters of the penalty term ($\alpha0$, $\alpha(t)$, etc.) and the starting guess for the field} define the final shape of the solution. A smart choice of {\color{black} parameters} can lead to a {\it better} optimal solution depending on the desired characteristic for the laser pulse (amplitude, shape, length) which in turn depend on the system under study and the available experimental set up. 
Our interest is to compare results obtained in vacuo and in PCM, exploring some possibilities for the {\color{black} penalty term} parameters and how they can affect the final results in the different environments {\color{black} (the role of the initial field guess is briefly discussed in Appendix \ref{sec:appB})}. 

\begin{figure}[h!]
\centering
  \includegraphics[width=1\linewidth]{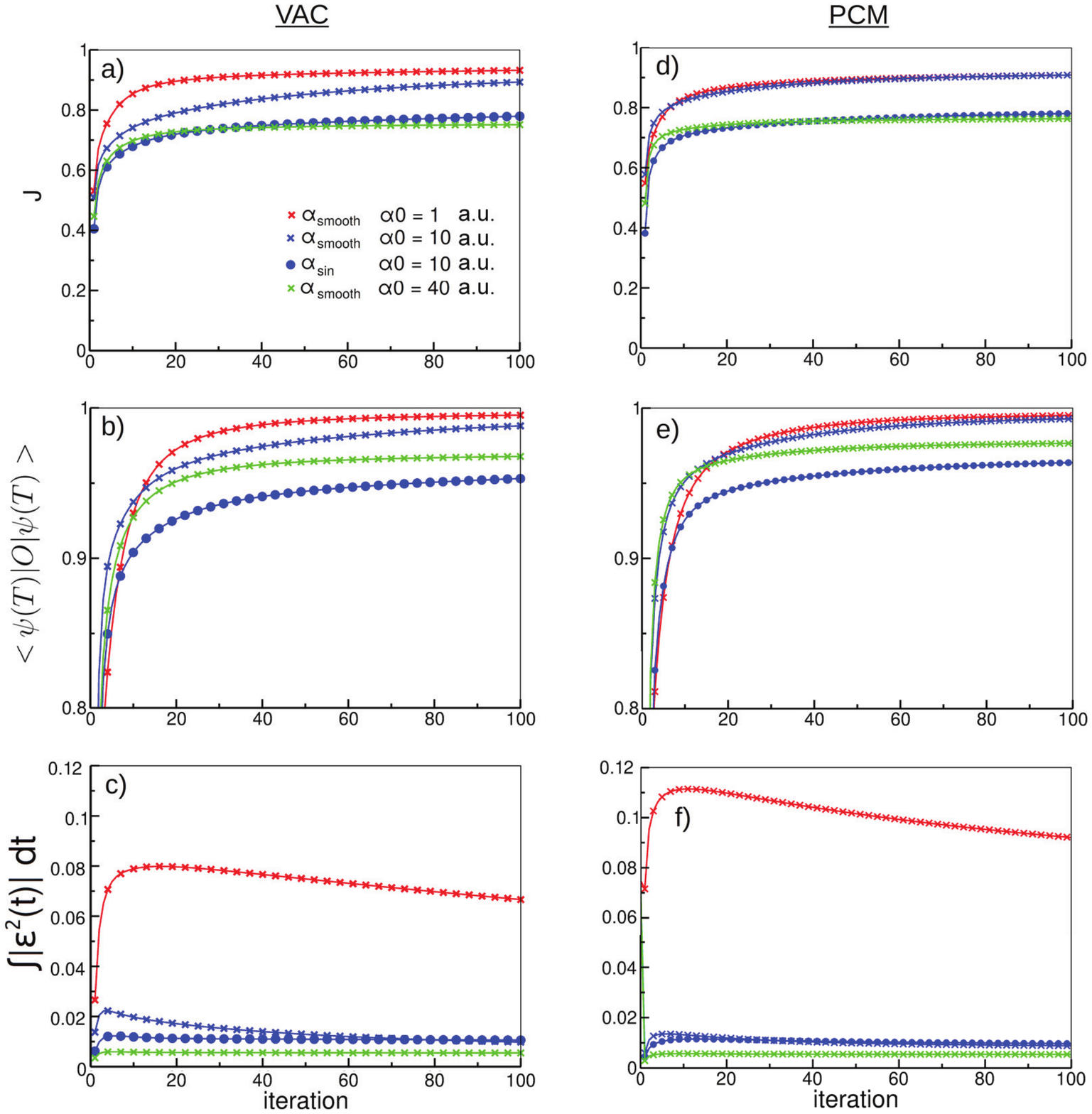}
\caption{Optimal control applied to  MQ in vacuo (left panel) and acetonitrile (right panel), with target state $|1\rangle$ (i.e., $\hat{O}=|1\rangle\langle 1|$ in vacuo and $\hat{O}=|1^{eq}\rangle\langle 1^{eq}|$ in solution). Plots of (a-d) J, (b-e) $\langle\psi(T) | \hat O |\psi(T)\rangle$ and (c-f) $\int_{0}^{T}$ $|\boldsymbol{\varepsilon}(t)|^{2}$ {\it dt} for different shapes of $\alpha(t)$ and different values of $\alpha0$. The starting field is $\boldsymbol{\varepsilon}^0_{0.01}$ = (0.01, 0.01, 0.01) a.u. Values are plotted one every two iterations}
\label{fig:quin}
\end{figure}

{\it Effect of the shape factor $\alpha(t)$.} Figure \ref{fig:quin} analyze the differences in the OC algorithm behaviour as obtained by the different choice of the shape field factor $\alpha(t)$ (see Eq. \ref{eq:alpha}). The three plots in the left panel refer to calculations performed in vacuo, while the right panel refers to calculations performed in acetonitrile implicit solvent that will be commented later in this work.\\
We here show calculations only with $\alpha(t) = \alpha_{smooth}(t)$ and $\alpha(t) = \alpha_{sin}(t)$, as the $\alpha_{const}$ shape gives practically the same results as $\alpha_{smooth}$, which only differs for a small penalty for high fields at the beginning and the end of the time interval, that accounts for the experimental turning on and turning off of the laser pulse. 

In Figure \ref{fig:quin} {\it left panel} it is shown how the $\alpha_{sin}$ shape is the one that performs worst in terms of reaching the target state. While for $\alpha0$=1 a.u. the overall performance ({\it J} value) is similar to the ones of of $\alpha(t) = \alpha_{smooth}(t)$, {\it J} becomes increasingly smaller with larger values of $\alpha0$, without a significant gain in the value of the field amplitude.\\
Even if this particular shape performs poorly, it is useful to be able to chose a specific shape for the optimized pulse, e. g. for experimental reasons. In this case the $\alpha_{sin}$ shape starts and ends to zero, similarly to a $\pi$-pulse shape. Other shapes of interest could be also implemented.\\
As expected, the integrals of the field are generally smaller with increasing values of $\alpha0$, with a consequence on the value of $\langle\psi(T) | \hat O |\psi(T)\rangle$. As we have already pointed out, in the case of $\alpha(t) = \alpha_{sin}(t)$ the loss in performance is particularly severe. 
We compare our results with the ones of Ref \cite{klamroth2006optimal} for optimal control on MQ for a 250 a.u. pulse. After 30 iteration they report achieving a population of the target state larger than 95\%. We use a different propagation algorithm for the wave function and a different starting field, but we obtain the same result with both $\alpha0$ $=$ 10 a.u. and $\alpha0$ $=$ 40 a.u. ($\alpha(t)= \alpha_{smooth}(t)$), and similar values for the field amplitude.  
The shape and frequency distribution of the optimal pulse are discussed in the following and compared with the ones in  Ref. \cite{klamroth2006optimal}.\\

{\it Effect of the solvent.} We focus now on the comparison with solvated MQ, implementing the theory described in Sec. \ref{sec:PCM}.
In the {\it right panel} of Fig. \ref{fig:quin}, { \it J}, $\langle\psi(T) | \hat O |\psi(T)\rangle$ and $\int_{0}^{T}$ $|\boldsymbol{\varepsilon}_M(t)|^{2}$ {\it dt} are plotted for MQ in acetonitrile solvent. \\
The target state for MQ in solution is the first excited state, as in vacuo (i.e., $\hat{O}=|1^{eq} \rangle \langle 1^{eq} |$); the solvent affects the state energy (see Fig. \ref{fig:quin-energy}) but for this molecule has a negligible  effect on its wavefunction (i.e., $|1^{eq}\rangle \approx |1^{fro}\rangle  \approx |1^{vac}\rangle$). This choice allows to study the effect of the solvent inclusion on a simple system, where the presence of the solvent only slightly modifies the electronic structure of the molecule, before moving to more complex scenarios. The optimal control calculations give very similar results to the in vacuo case, showing that the additional non-linear terms in the Hamiltonian of the solvated molecules are not changing the monotonic behavior and the overall performance of the algorithm. Comparing the magnitude of the fluency obtained in solution w.r.t. the in vacuo case, Fig. \ref{fig:quin}, we do not find a clear trend (depending on the optimization parameters, it may increase or decrease). This indirectly shows that the effect of the solvent is not only in the cavity field term (that would generically magnifies the Maxwell field), the time dependent reaction field has a role too. 

{\it Frequency analysis of the optimal laser field.} To better understand the characteristics of the optimal control field with respect to the system under study, and its dependence on the target state and on the starting optimal control parameters, we analyze the Fourier transform of the optimal control pulses obtained after 100 iterations with different starting conditions.\\

\begin{figure}[h!]
\centering
  \includegraphics[width=1\linewidth]{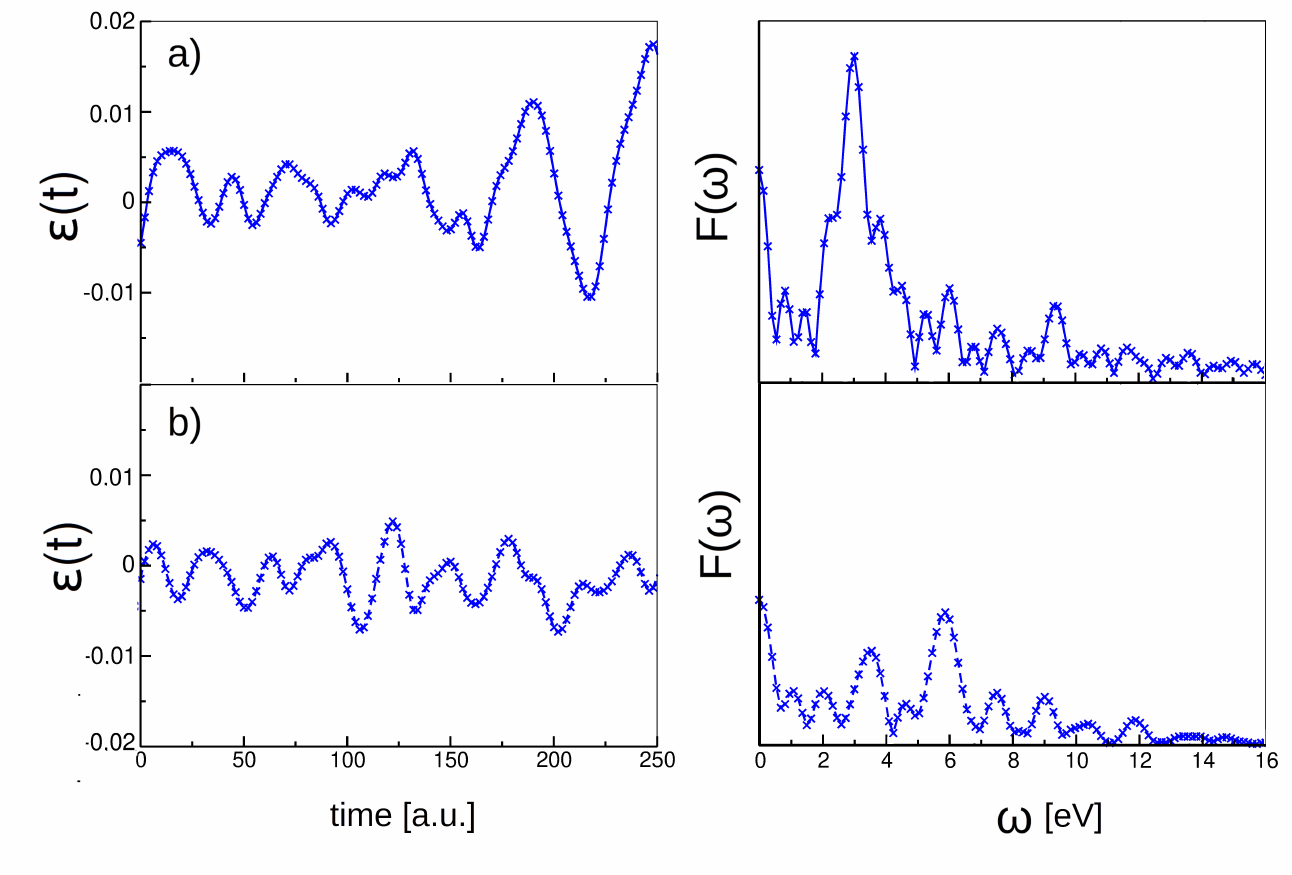}
\caption{MQ in vacuo: optimal pulse and frequency distribution for directions parallel (a) and perpendicular (b) to the 0-1 transition dipole moment direction, obtained with $\alpha(t)=\alpha_{smooth}$ and $\alpha0$ = 10 a.u..}
\label{fig:confronto}
\end{figure}

\begin{figure}[h!]
\centering
  \includegraphics[width=1\linewidth]{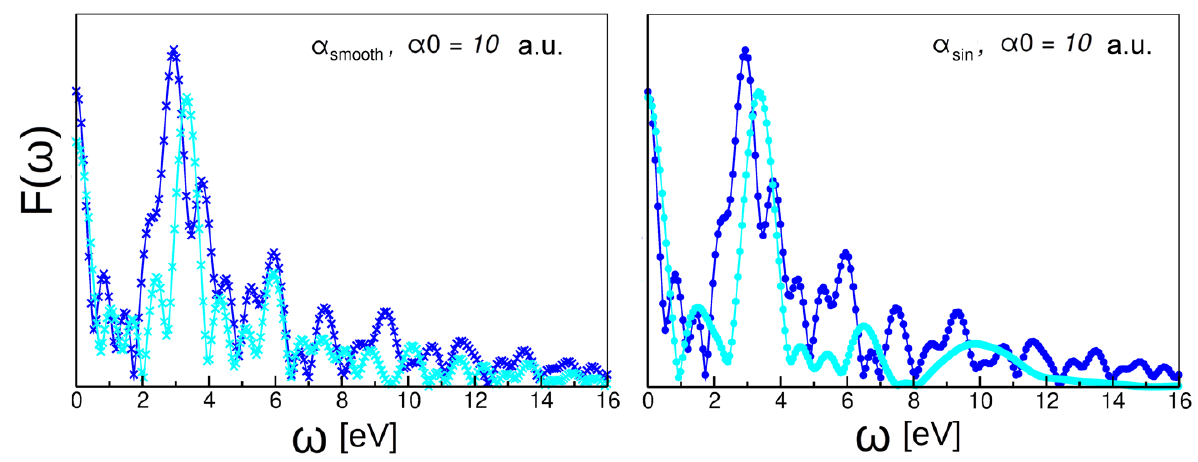}
\caption{Frequency distribution of optimal field in direction parallel to the 0-1 transition dipole moment for a) $\alpha(t)  = \alpha_{smooth}(t)$ and b) $\alpha(t) = \alpha_{sin}(t)$, with $\alpha0$ = 10 a.u. for in vacuo (blue) and solvated (cyan) systems.}
\label{fig:fourier}
\end{figure}

First, we compare results obtained with $\alpha(t)=\alpha_{smooth}$ and $\alpha0$=10 a.u. with the results in Ref. \cite{klamroth2006optimal} with the same parameters on the same molecule, with the aim of validating our numerical implementation. In Figure \ref{fig:confronto} a) the field and Fourier transform on the direction parallel to the transition dipole moment are plotted, while in b) the direction is perpendicular to the dipole. On the parallel direction the main peak of the Fourier transform corresponds to the resonant frequency for the 0-1 transition, at  3.118 eV. A $\pi$-pulse with this frequency would cause a population inversion in a two level system but, as already mentioned, in Ref. \cite{klamroth2006optimal} is discussed how this approximation does not work for the system under study for short pulses (250 a.u.) and how an optimized pulse is needed instead. Nevertheless a significant component of the field corresponds to the resonance frequency. Concerning the perpendicular component of the Fourier transform, there is a peak corresponding to a static component at nearly zero frequency and two others peaks around 3.5 eV and 5.8 eV.  These additional frequencies allow the optimized pulse to populate the first excited state, discouraging further excitation and de-excitation to competitive states. These results are in complete agreement with the ones obtained in Ref. \cite{klamroth2006optimal}.\\
We then compare the results obtained with different $\alpha(t)$ shapes in vacuo and in implicit solvent. Only the component parallel to the transition dipole moment is shown for clarity (Fig. \ref{fig:fourier} a,b)).

The two $\alpha(t)$ shapes give very similar results in terms of frequency, both in vacuo and in acetonitrile, with the main peak corresponding to the resonance energy for the 0-1 transition. In implicit solvent this frequency is at $\omega$= 3.565 eV, blue shifted with respect to gas-phase, as expected from free energy values in Fig. \ref{fig:quin-energy}, and a second peak at $\omega \simeq 6$ eV is visible for both shapes of $\alpha(t)$.\\
A deeper analysis of the populations behaviour in time under the influence of the optimal field, and of the fluxes between excited states, it will be carried out in a future paper. 

\subsubsection{LiCN}
\begin{figure}[h!]
\centering
 \includegraphics[width=0.7\linewidth]{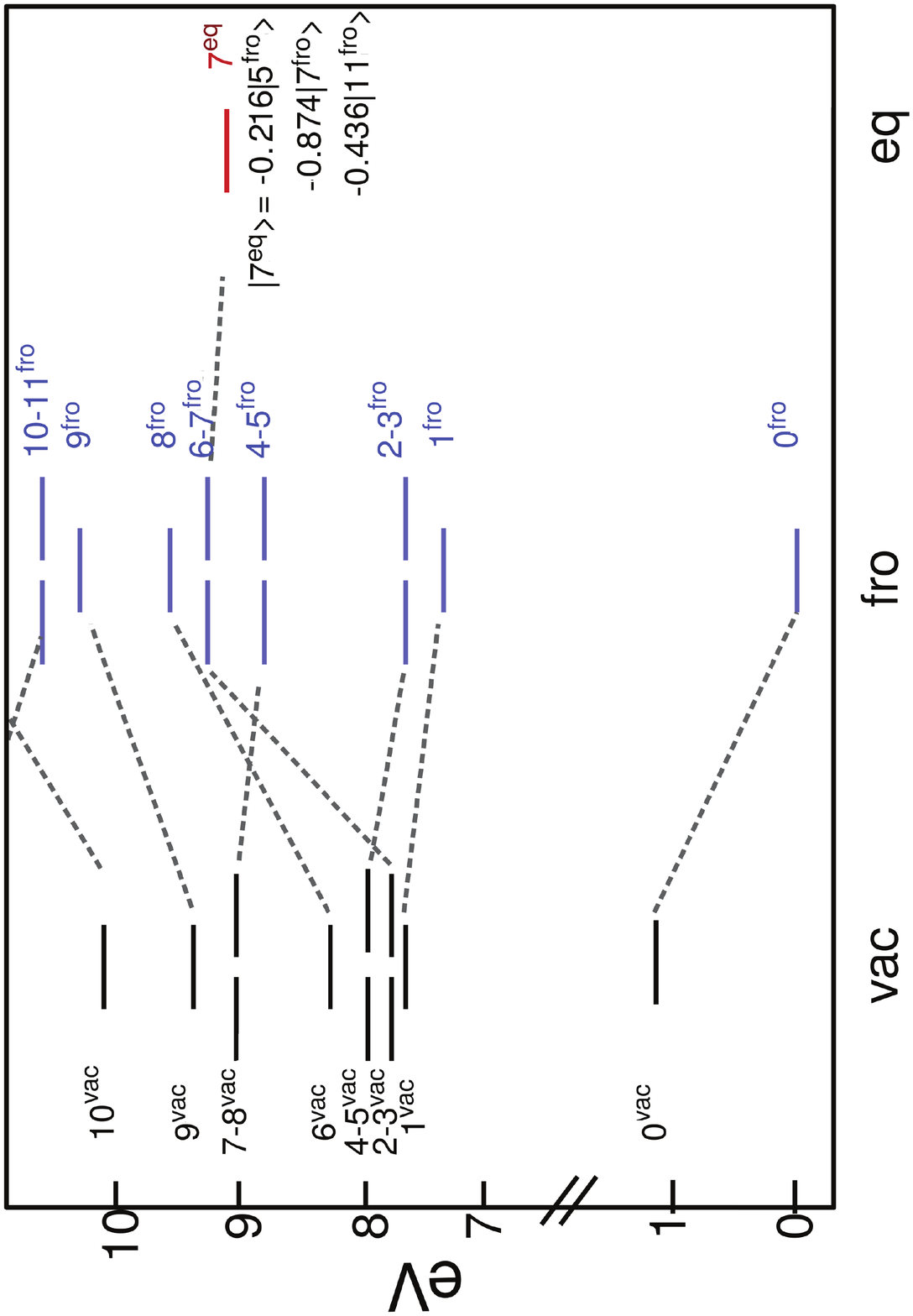}
\caption{Energies  levels  of  the  ground and the lowest excited states for LiCN in-vacuo (black), within the frozen solvent approximation (blue), and  after the equilibration of the solvent with the excitation.  All the values refer to the equilibrium free energy of the ground state in solution which is therefore set to zero.}
\label{fig:LiCN_energies}
\end{figure}

LiCN was chosen due to the characteristics of its electronic structure and to how it is modified by the presence of the solvent. In Figure \ref{fig:molecules} b) we show the direction and value of the transition dipole of the 0-2 transition (0-7 in PCM), while to better understand the effect of the solvent on the position of the different levels in the two environments and to guide the choice of the target state, we report in Fig. \ref{fig:LiCN_energies} a correlation plot showing the energies of the lowest states in gas-phase and their non-equilibrium free-energy within the frozen solvent approximation. From the figure it is apparent that some states are very sensitive to the solvent effects,\cite{pipolo2017equation} and for them we also expect some non-negligible change in the wavefunction.

For LiCN in vacuo we chose as target state the second excited state (the same chosen in a previous study,\cite{klamroth2006optimal} that we here extend to the solvated case). Figure \ref{fig:LiCN_energies} shows that it corresponds to the sixth excited state in solution (in the frozen solvent approximation). Moreover we have to take into account that solvent equilibration may change the states. By solving Eq.(\ref{eq:h_state}) using the self consistent algorithm described in Sec. \ref{sec:pcm_rev}, we find in fact how $|7^{eq} \rangle$ is written in terms of frozen solvent states after the solvent is equilibrated with the molecule ($|7^{eq}\rangle$ = $-0.216|5^{fro}\rangle$$-0.874|7^{fro}\rangle $$-0.436|11^{fro}\rangle$). This choice allows to test the algorithm in PCM for a complex target state, written as a linear combination of frozen solvent ones (see Sec. \ref{sec:pcm_rev}). 

\begin{figure}[h!]
\centering
 \includegraphics[width=1\linewidth]{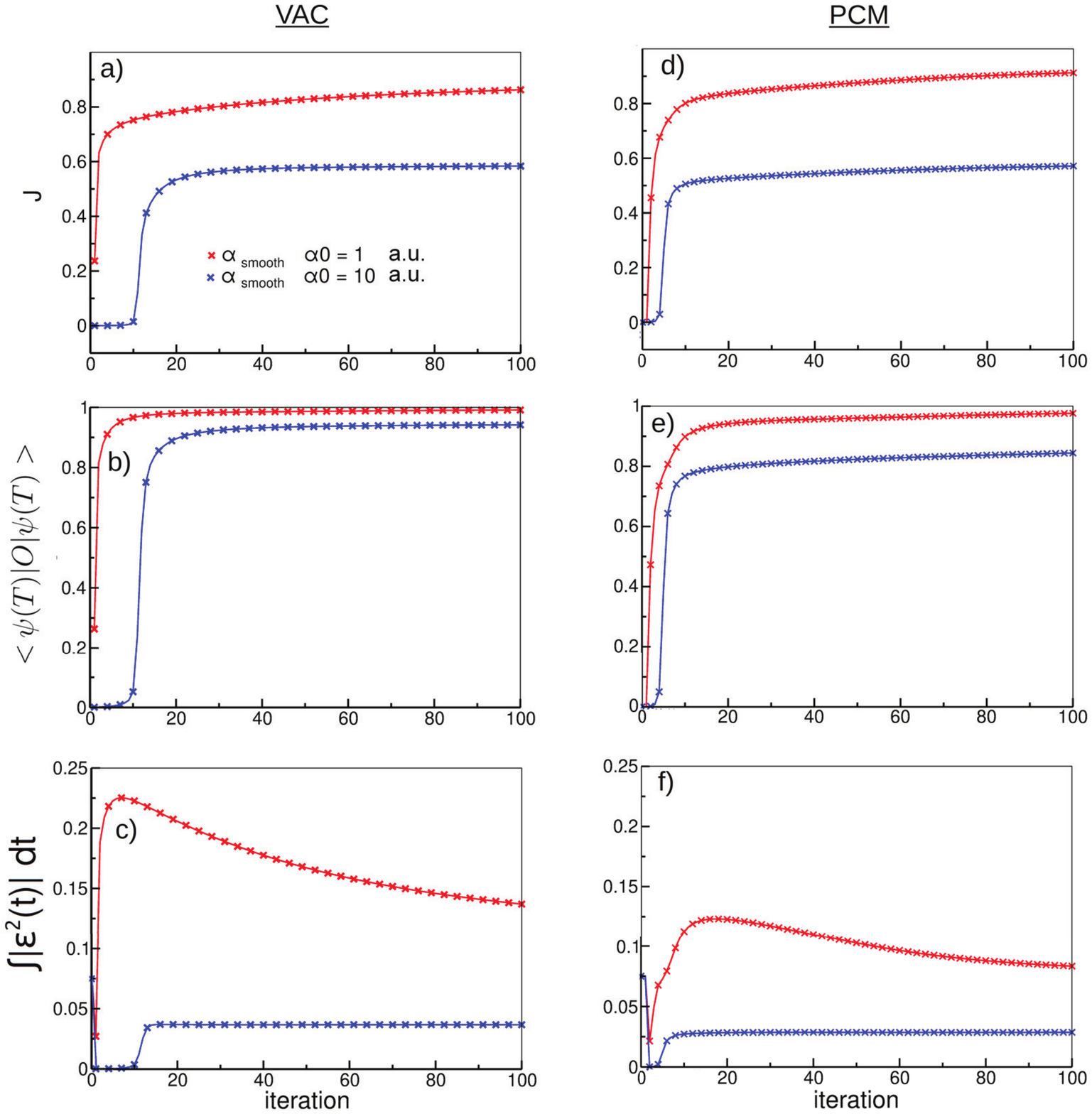}
\caption{Optimal control applied to LiCN molecule in vacuo, with target state $|2\rangle$ (left panel) and to LiCN molecule in acetonitrile, with target state $|7^{eq}\rangle$ (right panel). (a-d) J, (b-e) $\langle\psi(T)| \hat O |\psi(T)\rangle$ and (c-f) $\int_{0}^{T}$ $|\boldsymbol{\varepsilon}(t)|^{2}$ {\it dt} for different shapes of $\alpha(t)$ and different values of $\alpha0$. The starting field is $\boldsymbol{\varepsilon}^0_{0.01}$ = (0.01, 0.01, 0.01) a.u.. Values are plotted every second iteration}
\label{fig:LiCN}
\end{figure}

{\it LiCN: QOCT in vacuo.} Results for LiCN in vacuo (Fig. \ref{fig:LiCN} {\it left panel}) are similar to the ones obtained for MQ, with some differences due to the different nature of the molecule. For the sake of clarity, we restricted our study to $\alpha(t) = \alpha_{smooth}(t)$, that was proposed previously\cite{klamroth2006optimal} and gave good results for MQ. On the contrary, simpler fields with plain shapes as the $\alpha_{smooth}$ one seem to be more suitable as a starting point to satisfy the problem requirements. LiCN 0-2 transition has a smaller value of the transition dipole with respect to MQ 0-1 excitation, as a consequence an effective choice of $\alpha(t)$ is even more important.\\ 
In the calculation with $\alpha0$ = 10 a.u. the optimal control algorithm is slower to find the desired solution, with very low values of $\langle\psi(T)| \hat O |\psi(T)\rangle$ in the first iterations (notice that to show this behaviour in Fig. \ref{fig:LiCN} the two plots for $\langle\psi(T)| \hat O |\psi(T)\rangle$ are plotted with y axes starting from 0, differently with respect to Fig. \ref{fig:quin})
Values for $\int_{0}^{T}$ $|\boldsymbol{\varepsilon}(t)|^{2}$ {\it dt} are larger with respect to MQ, showing how stronger fields are needed to populate the desired excited state in LiCN. On the contrary, the final value of $\langle\psi(T)| \hat O |\psi(T)\rangle$ in vacuo is very similar for MQ and LiCN. As a consequence, {\it J} in LiCN is smaller than for MQ, as it accounts for the larger value of the field integral.\\

{\it LiCN: QOCT-PCM in solution.} Comparing the results obtained for LiCN in vacuo (Fig.\ref{fig:LiCN} {\it left panel}) and in PCM (Fig.\ref{fig:LiCN} {\it right panel}) in terms of performances of the algorithm, the results are very similar (independently from the values of $\alpha0$) in terms of final {\it J}, target state population and optimized field, showing that in solution it is possible to get the same performance as in vacuo. 
 Hence, at least for the examples shown here, the additional non-linear terms (reaction field terms) of the Hamiltonian for the solvated system (see Eq. (\ref{eq:sh_pcm})) do not modify the performance of the QOCT algorithm. In addition, we also remark that although in the case of the solvated molecules (i.e. in PCM) the target state for LiCN ($|7^{eq}\rangle$) is a linear combination of the eigenstates basis set (three of them dominate, see Fig.(\ref{fig:LiCN_energies})), the performances of the QOCT algorithm in vacuo are again retained.

Eventually from our data we can say that optimal control performance (in terms of {\it J } and on how close one can go to a target excited state) is very system and state dependent and a careful choice of the optimal control parameters and starting field can help to achieve better results.\\

\section{Conclusions}
In this study, we extended QOCT to the case of solvated molecules. In particular, to account for the solvent, we used a PCM approach and we included into the effective Hamiltonian of the molecule both the term describing the solvent polarization induced by the time dependent charge distribution of the solute (reaction field) and the interaction with the electric field associated with the incoming light pulse as modified by the solute cavity (cavity field effects). The present work extends thus the recently developed time dependent PCM theory to the case of QOCT.\cite{Pipolo2014,Corni2014,pipolo2017equation}

The presence of the dielectric medium modifies the optical and electronic properties of the molecular system as the electronic dynamics of the molecule is coupled with the one of the solvent. In particular, the resulting objective functional $J$ for a quantum solute in a dielectric solvent contains a non-linear term, a feature found before in a different context \cite{sklarz2002loading} (a local non-linearity there, a global non-linearity in our case) that leads to QOCT evolution equations with additional terms with respect to the in vacuo case.


We applied the newly developed QOCT-PCM approach to two molecules in solution, LiCN and MQ, that were investigated before either in the contest of TD-PCM\cite{Pipolo2014,Corni2014,pipolo2017equation} or QOCT.\cite{klamroth2006optimal} From our data we found that the inclusion of PCM terms into the Hamiltonian do not seem to substantially modify the performance of the optimal control algorithm. Indeed in our calculaions it was possible to achieve similar final target state populations in solvent as in vacuo, and through a comparable number of iterations. The main difference in the amplitude of the optimal field which, depending on the system, is smaller or larger in PCM than in vacuo, indirectly shows that the effect of the solvent is both in the cavity field term and in the time dependent reaction field.
For the solution case, we also pointed out that the target excited state can be chosen to be that equilibrated with the solvent. Such state can be represented as a linear combination of non-equilibrium (frozen solvent) excited states.

In conclusion, the theory and the implementation described here provide the tools to include in a computationally affordable way the effect of a solvent in the design, by QOCT, of light pulses able to take a solute in a desired state. 

\appendix
{\color{black}
\section{{\it J} functional derivation}}
\subsection{In vacuo system}
To compute the stationary points of {\it J}, i.e. $\partial J=0$, 
it is convenient to use integration by part in Eq. (\ref{eq:J}) to get:
\begin{equation}\label{eq:J_parts}\begin{split}
J = &\langle\psi(T)|\hat O| \psi(T)\rangle - \int_{0}^{T} \alpha(t)|\boldsymbol{\varepsilon}(t)|^{2} dt \\
&- \langle\chi(t) | \psi(t)\rangle|_0^T 
+\Bigg [ \int_{0}^{T} \langle \frac{\partial}{\partial t} \chi(t)| \psi(t)\rangle - \langle \chi(t) | \: {\it i} [ \hat H_0 - \boldsymbol{\varepsilon}(t)\boldsymbol{\hat{{\mu}}}] | \psi(t) \rangle dt \Bigg ] \\
&
- \Bigg [ \int_{0}^{T} \langle \left[ \frac{\partial}{\partial t} + {\it i} ( \hat H_0 - \boldsymbol{\varepsilon}(t)\boldsymbol{\hat{{\mu}})\right]} \psi(t) | \chi(t) \rangle dt \Bigg ] \end{split}
\end{equation}

To compute the stationary points of {\it J} we have now to differentiate with respect to $\psi(t)$, $\chi(t)$ and $\boldsymbol{\varepsilon}(t)$:

\begin{eqnarray}\label{eq:wf_diff}
\nonumber \partial_{|\psi\rangle}J &=& \langle\psi(T)|\hat O| \partial\psi(T)\rangle -\langle\chi(T) | \partial \psi(T)\rangle+\\ &&\int_{0}^{T} \langle \frac{\partial}{\partial t}\chi(t)|\partial\psi(t)\rangle - \langle\chi(t)| \:{\it i} [ \hat H_0 - \boldsymbol{\varepsilon}(t)\boldsymbol{\hat{{\mu}}}] | \partial \psi(t) \rangle dt
\end{eqnarray}

\begin{equation}\label{eq:chi_diff}
\partial_{|\chi\rangle}J = - \Bigg [ \int_{0}^{T} \langle \frac{\partial}{\partial t} \psi(t)|\partial \chi(t)\rangle + \langle {\it i} [ \hat H_0 - \boldsymbol{\varepsilon}(t)\boldsymbol{\hat{{\mu}}}] \psi(t)| \partial \chi(t) \rangle dt \Bigg ]
\end{equation}

\begin{equation}\label{eq:field_diff}
\partial_{\varepsilon}J = - \Bigg[ \int_{0}^{T} \left( -2Im \langle \chi(t)|\boldsymbol{\hat{{\mu}}} | \psi(t) \rangle - 2 \alpha(t) \boldsymbol{\varepsilon}(t) \partial \boldsymbol{\varepsilon}(t) \right) dt \Bigg]
\end{equation}

\subsection{Solvated system}
\label{sec:a2}
Similarly to what was done in vacuo, we differentiate Eq. \ref{eq:J_PCM} with respect to $\psi(t)$, $\chi(t)$ and $\boldsymbol{\varepsilon}(t)$:

\begin{equation}\label{eq:wf_diff_pcm}\begin{split}
\partial_{|\psi\rangle}J^{PCM} = & \langle\psi(T)|\hat O| \partial\psi(T)\rangle -\langle\chi(T) | \partial \psi(T)\rangle + \\
& \int_{0}^{T} \Bigg( \langle \frac{\partial}{\partial t} \chi(t) | \partial \psi(t)\rangle - \langle \chi(t)  |\: {\it i} [ \hat H^{0} + (\langle\psi(t) | {\bf \hat q}_d|\psi(t)\rangle +{\bf q}_{in})\cdot {\bf \hat V} -\boldsymbol{\varepsilon}_M(t)\boldsymbol{\hat{\bar{\mu}}}]| \partial \psi(t) \rangle \\
& - \langle \chi(t)|\:{\it i} {\bf \hat V} | \psi(t) \rangle \cdot \langle \psi(t)|{\bf \hat q}_d|\partial \psi(t)\rangle - \langle {\it i} {\bf \hat V} \psi(t)|\chi(t)\rangle \cdot \langle \psi(t)|{\bf \hat q}_d|\partial \psi(t)\rangle \Bigg) dt 
\end{split}
\end{equation}

\begin{equation}\label{eq:chi_diff_pcm}
\partial_{|\chi\rangle}J^{PCM} = -\Bigg[ \int_{0}^{T} \langle \frac{\partial}{\partial t} \psi(t)|\partial \chi(t)\rangle + \langle \: {\it i} [ \hat H^{0} + (\langle\psi(t)|{\bf \hat q}_d | \psi(t) \rangle+{\bf q}_{in}) \cdot {\bf \hat V} - \boldsymbol{\varepsilon}_M(t)\boldsymbol{\hat{\bar{\mu}}}] \psi(t)| \partial \chi(t) \rangle dt \Bigg ]
\end{equation}

\begin{equation}\label{eq:field_diff_pcm}
\partial_{\varepsilon_M}J^{PCM} = - \Bigg[\int_{0}^{T} \langle-2Im \langle \chi(t)|\boldsymbol{\hat{\bar{\mu}}} | \psi(t) \rangle - 2 \alpha(t) \boldsymbol{\varepsilon}(t) \partial \boldsymbol{\varepsilon}_M(t)\Bigg]
\end{equation}

The main difference in the result with respect to vacuo is a consequence of the $\langle\psi(t)|{\bf \hat q} | \psi(t) \rangle$ term, which generates extra terms in Eq. (\ref{eq:wf_diff_pcm}) and, as a consequence, in the backward propagation term of the optimal control problem in Eq. (\ref{eq:chi_pcm}).

\begin{figure}[h!]
\centering
 \includegraphics[width=0.5\linewidth]{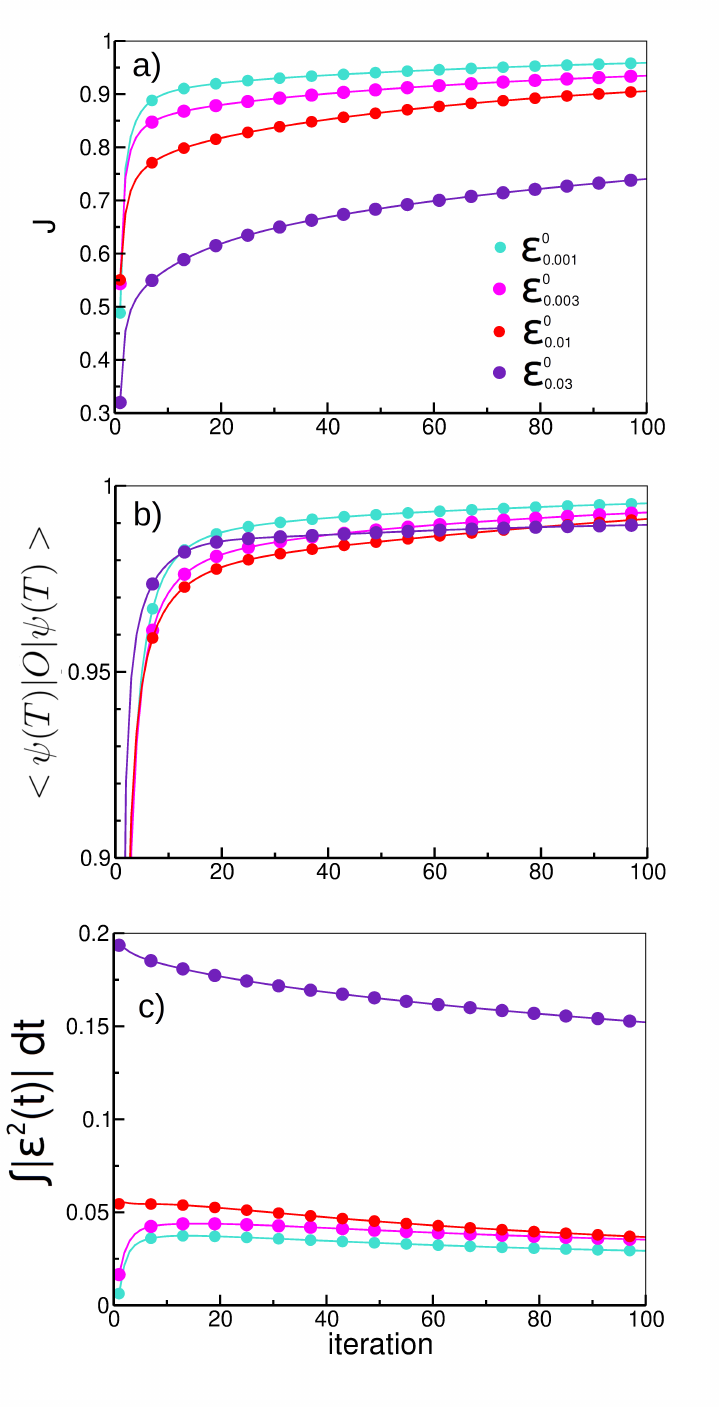}
\caption{Optimal control applied to MQ in vacuo, with target state $|1\rangle$. Plots of (a) J, (b) $\langle\psi(T) \hat O |\psi(T)\rangle$ and (c) fluency $\int_{0}^{T}$ $|\boldsymbol{\varepsilon}(t)|^{2}$ {\it dt} with respect the number of iterations with different starting fields (given in atomic units), $\alpha_{sin}(t)$ is used with $\alpha0$=1 a.u.. Values are plotted every fifth iterations.}
\label{fig:quin_vac_field}
\end{figure}

{\color{black}
\section{Optimal control of MQ in vacuo with different starting electric fields}}
\label{sec:appB}

{\color{black}In this Appendix we want to to briefly discuss the choice of the guess field at the beginning of the OC iterations. To this purpose we applied the optimal control problem to MQ in vacuo with  different starting fields: $\boldsymbol{\varepsilon}^0_{0.001}$ = (0.001, 0.001, 0.001) a.u., $\boldsymbol{\varepsilon}^0_{0.003}$ = (0.003, 0.003, 0.003) a.u.,  $\boldsymbol{\varepsilon}^0_{0.01}$=(0.01, 0.01, 0.01) a.u. and $\boldsymbol{\varepsilon}^0_{0.03}$=(0.03,0.03,0.03) a.u. (Fig. \ref{fig:quin_vac_field})\\
 In the case studied, the value of $\langle\psi(T)|\hat O|\psi(T)\rangle$ after 100 iterations is very similar in the four cases, while the fields behave quite differently: the three smaller fields converge to the same final value of amplitude, while the larger field reaches the same value of $\langle\psi(T)|\hat O|\psi(T)\rangle$, but its final amplitude is much larger. In practice it would need a large amount of iterations to converge to a value similar to the one of the other three cases. In this particular case $\boldsymbol{\varepsilon}^0_{0.001}$ is the starting pulse which guarantees the better performance, which means the smaller field with the same final value of $\langle\psi(T)|\hat O|\psi(T)\rangle$. Nevertheless, depending on the values of the transition dipoles and the orientation of the molecule, it can happen that the final value of $\hat O |\psi(T)\rangle$ after the first iteration(s) is too small to provide the {\it information} needed by the algorithm in order to improve the field at the next step (i. e., a starting zero field will not allow the algorithm to work). In such cases, a larger field must be used, with the consequence that is simply a larger amplitude of the optimized field (Fig. \ref{fig:quin_vac_field}). For this reason we have used $\boldsymbol{\varepsilon}^0_{0.01}$=(0.01, 0.01, 0.01) a.u. as starting field for the bulk of our calculations.} \\

\begin{acknowledgments}
 Funding from the EU H2020 ERC under the grant ERC-CoG-681285 TAME-Plasmons is gratefully acknowledged. Computational work has been partially carried out on the C3P (Computational Chemistry Community in Padua) HPC facility of the Department of Chemical Sciences of the University of Padua.
\end{acknowledgments}

\bibliography{OC_nocolor}

\begin{thebibliography}{49}
\expandafter\ifx\csname natexlab\endcsname\relax\def\natexlab#1{#1}\fi
\expandafter\ifx\csname bibnamefont\endcsname\relax
  \def\bibnamefont#1{#1}\fi
\expandafter\ifx\csname bibfnamefont\endcsname\relax
  \def\bibfnamefont#1{#1}\fi
\expandafter\ifx\csname citenamefont\endcsname\relax
  \def\citenamefont#1{#1}\fi
\expandafter\ifx\csname url\endcsname\relax
  \def\url#1{\texttt{#1}}\fi
\expandafter\ifx\csname urlprefix\endcsname\relax\def\urlprefix{URL }\fi
\providecommand{\bibinfo}[2]{#2}
\providecommand{\eprint}[2][]{\url{#2}}

\bibitem[{\citenamefont{Rabitz}(2002)}]{Rabitz2002}
\bibinfo{author}{\bibfnamefont{H.}~\bibnamefont{Rabitz}}, in
  \emph{\bibinfo{booktitle}{Encyclopedia of Computational Chemistry}}
  (\bibinfo{publisher}{John Wiley {\&} Sons, Ltd},
  \bibinfo{address}{Chichester, UK}, \bibinfo{year}{2002}).

\bibitem[{\citenamefont{Tannor and Rice}(1985)}]{Tannor1985}
\bibinfo{author}{\bibfnamefont{D.~J.} \bibnamefont{Tannor}} \bibnamefont{and}
  \bibinfo{author}{\bibfnamefont{S.~A.} \bibnamefont{Rice}},
  \bibinfo{journal}{J. Chem. Phys.} \textbf{\bibinfo{volume}{83}},
  \bibinfo{pages}{5013} (\bibinfo{year}{1985}).

\bibitem[{\citenamefont{Peirce et~al.}(1988)\citenamefont{Peirce, Dahleh, and
  Rabitz}}]{Peirce1988}
\bibinfo{author}{\bibfnamefont{A.~P.} \bibnamefont{Peirce}},
  \bibinfo{author}{\bibfnamefont{M.~A.} \bibnamefont{Dahleh}},
  \bibnamefont{and} \bibinfo{author}{\bibfnamefont{H.}~\bibnamefont{Rabitz}},
  \bibinfo{journal}{Phys. Rev. A} \textbf{\bibinfo{volume}{37}},
  \bibinfo{pages}{4950} (\bibinfo{year}{1988}).

\bibitem[{\citenamefont{Kosloff et~al.}(1989)\citenamefont{Kosloff, Rice,
  Gaspard, Tersigni, and Tannor}}]{kosloff1989wavepacket}
\bibinfo{author}{\bibfnamefont{R.}~\bibnamefont{Kosloff}},
  \bibinfo{author}{\bibfnamefont{S.~A.} \bibnamefont{Rice}},
  \bibinfo{author}{\bibfnamefont{P.}~\bibnamefont{Gaspard}},
  \bibinfo{author}{\bibfnamefont{S.}~\bibnamefont{Tersigni}}, \bibnamefont{and}
  \bibinfo{author}{\bibfnamefont{D.}~\bibnamefont{Tannor}},
  \bibinfo{journal}{Chem. Phys.} \textbf{\bibinfo{volume}{139}},
  \bibinfo{pages}{201} (\bibinfo{year}{1989}).

\bibitem[{\citenamefont{Keefer and de~Vivie-Riedle}(2018)}]{keefer2018pathways}
\bibinfo{author}{\bibfnamefont{D.}~\bibnamefont{Keefer}} \bibnamefont{and}
  \bibinfo{author}{\bibfnamefont{R.}~\bibnamefont{de~Vivie-Riedle}},
  \bibinfo{journal}{Acc. Chem. Res.} \textbf{\bibinfo{volume}{51}},
  \bibinfo{pages}{2279} (\bibinfo{year}{2018}).

\bibitem[{\citenamefont{Deffner}({2014})}]{deffner2014optimal}
\bibinfo{author}{\bibfnamefont{S.}~\bibnamefont{Deffner}}, \bibinfo{journal}{{J
  Phys. B}} \textbf{\bibinfo{volume}{{47}}}, \bibinfo{pages}{{145502}}
  (\bibinfo{year}{{2014}}).

\bibitem[{\citenamefont{Allen et~al.}({2017})\citenamefont{Allen, Kosut, Joo,
  Leek, and Ginossar}}]{allen2017optimal}
\bibinfo{author}{\bibfnamefont{J.~L.} \bibnamefont{Allen}},
  \bibinfo{author}{\bibfnamefont{R.}~\bibnamefont{Kosut}},
  \bibinfo{author}{\bibfnamefont{J.}~\bibnamefont{Joo}},
  \bibinfo{author}{\bibfnamefont{P.}~\bibnamefont{Leek}}, \bibnamefont{and}
  \bibinfo{author}{\bibfnamefont{E.}~\bibnamefont{Ginossar}},
  \bibinfo{journal}{{Phys. Rev. A}} \textbf{\bibinfo{volume}{{95}}},
  \bibinfo{pages}{{042325}} (\bibinfo{year}{{2017}}).

\bibitem[{\citenamefont{Castro et~al.}({2019})\citenamefont{Castro, Appel, and
  Rubio}}]{castro2019optimal}
\bibinfo{author}{\bibfnamefont{A.}~\bibnamefont{Castro}},
  \bibinfo{author}{\bibfnamefont{H.}~\bibnamefont{Appel}}, \bibnamefont{and}
  \bibinfo{author}{\bibfnamefont{A.}~\bibnamefont{Rubio}},
  \bibinfo{journal}{{EPJ B}} \textbf{\bibinfo{volume}{{92}}},
  \bibinfo{pages}{{223}} (\bibinfo{year}{{2019}}).

\bibitem[{\citenamefont{Henriksen}(2002)}]{henriksen2002laser}
\bibinfo{author}{\bibfnamefont{N.~E.} \bibnamefont{Henriksen}},
  \bibinfo{journal}{Chem. Soc. Rev.} \textbf{\bibinfo{volume}{31}},
  \bibinfo{pages}{37} (\bibinfo{year}{2002}).

\bibitem[{\citenamefont{Ma et~al.}(2002)\citenamefont{Ma, Tafti, and
  Braatz}}]{ma2002optimal}
\bibinfo{author}{\bibfnamefont{D.~L.} \bibnamefont{Ma}},
  \bibinfo{author}{\bibfnamefont{D.~K.} \bibnamefont{Tafti}}, \bibnamefont{and}
  \bibinfo{author}{\bibfnamefont{R.~D.} \bibnamefont{Braatz}},
  \bibinfo{journal}{Comp. \& Chem. Eng.} \textbf{\bibinfo{volume}{26}},
  \bibinfo{pages}{1103} (\bibinfo{year}{2002}).

\bibitem[{\citenamefont{Kehlet et~al.}(2004)\citenamefont{Kehlet, Sivertsen,
  Bjerring, Reiss, Khaneja, Glaser, and Nielsen}}]{kehlet2004improving}
\bibinfo{author}{\bibfnamefont{C.~T.} \bibnamefont{Kehlet}},
  \bibinfo{author}{\bibfnamefont{A.~C.} \bibnamefont{Sivertsen}},
  \bibinfo{author}{\bibfnamefont{M.}~\bibnamefont{Bjerring}},
  \bibinfo{author}{\bibfnamefont{T.~O.} \bibnamefont{Reiss}},
  \bibinfo{author}{\bibfnamefont{N.}~\bibnamefont{Khaneja}},
  \bibinfo{author}{\bibfnamefont{S.~J.} \bibnamefont{Glaser}},
  \bibnamefont{and} \bibinfo{author}{\bibfnamefont{N.~C.}
  \bibnamefont{Nielsen}}, \bibinfo{journal}{J. Am. Chem. Soc.}
  \textbf{\bibinfo{volume}{126}}, \bibinfo{pages}{10202}
  (\bibinfo{year}{2004}).

\bibitem[{\citenamefont{Kehlet et~al.}(2007)\citenamefont{Kehlet, Bjerring,
  Sivertsen, Kristensen, Enghild, Glaser, Khaneja, and
  Nielsen}}]{kehlet2007optimal}
\bibinfo{author}{\bibfnamefont{C.}~\bibnamefont{Kehlet}},
  \bibinfo{author}{\bibfnamefont{M.}~\bibnamefont{Bjerring}},
  \bibinfo{author}{\bibfnamefont{A.~C.} \bibnamefont{Sivertsen}},
  \bibinfo{author}{\bibfnamefont{T.}~\bibnamefont{Kristensen}},
  \bibinfo{author}{\bibfnamefont{J.~J.} \bibnamefont{Enghild}},
  \bibinfo{author}{\bibfnamefont{S.~J.} \bibnamefont{Glaser}},
  \bibinfo{author}{\bibfnamefont{N.}~\bibnamefont{Khaneja}}, \bibnamefont{and}
  \bibinfo{author}{\bibfnamefont{N.~C.} \bibnamefont{Nielsen}},
  \bibinfo{journal}{J. Magn. Reson.} \textbf{\bibinfo{volume}{188}},
  \bibinfo{pages}{216} (\bibinfo{year}{2007}).

\bibitem[{\citenamefont{Rabitz and Zhu}(2000)}]{rabitz2000optimal}
\bibinfo{author}{\bibfnamefont{H.}~\bibnamefont{Rabitz}} \bibnamefont{and}
  \bibinfo{author}{\bibfnamefont{W.}~\bibnamefont{Zhu}}, \bibinfo{journal}{Acc.
  Chem. Res.} \textbf{\bibinfo{volume}{33}}, \bibinfo{pages}{572}
  (\bibinfo{year}{2000}).

\bibitem[{\citenamefont{Geremia and Rabitz}(2002)}]{geremia2002optimal}
\bibinfo{author}{\bibfnamefont{J.}~\bibnamefont{Geremia}} \bibnamefont{and}
  \bibinfo{author}{\bibfnamefont{H.}~\bibnamefont{Rabitz}},
  \bibinfo{journal}{Phys. Rev. Lett.} \textbf{\bibinfo{volume}{89}},
  \bibinfo{pages}{263902} (\bibinfo{year}{2002}).

\bibitem[{\citenamefont{Daniel et~al.}(2003)\citenamefont{Daniel, Full,
  Gonz{\'a}lez, Lupulescu, Manz, Merli, Vajda, and
  W{\"o}ste}}]{daniel2003deciphering}
\bibinfo{author}{\bibfnamefont{C.}~\bibnamefont{Daniel}},
  \bibinfo{author}{\bibfnamefont{J.}~\bibnamefont{Full}},
  \bibinfo{author}{\bibfnamefont{L.}~\bibnamefont{Gonz{\'a}lez}},
  \bibinfo{author}{\bibfnamefont{C.}~\bibnamefont{Lupulescu}},
  \bibinfo{author}{\bibfnamefont{J.}~\bibnamefont{Manz}},
  \bibinfo{author}{\bibfnamefont{A.}~\bibnamefont{Merli}},
  \bibinfo{author}{\bibfnamefont{{\v{S}}.}~\bibnamefont{Vajda}},
  \bibnamefont{and}
  \bibinfo{author}{\bibfnamefont{L.}~\bibnamefont{W{\"o}ste}},
  \bibinfo{journal}{Science} \textbf{\bibinfo{volume}{299}},
  \bibinfo{pages}{536} (\bibinfo{year}{2003}).

\bibitem[{\citenamefont{Accanto et~al.}(2017)\citenamefont{Accanto, De~Roque,
  Galvan-Sosa, Christodoulou, Moreels, and Van~Hulst}}]{accanto2017rapid}
\bibinfo{author}{\bibfnamefont{N.}~\bibnamefont{Accanto}},
  \bibinfo{author}{\bibfnamefont{P.~M.} \bibnamefont{De~Roque}},
  \bibinfo{author}{\bibfnamefont{M.}~\bibnamefont{Galvan-Sosa}},
  \bibinfo{author}{\bibfnamefont{S.}~\bibnamefont{Christodoulou}},
  \bibinfo{author}{\bibfnamefont{I.}~\bibnamefont{Moreels}}, \bibnamefont{and}
  \bibinfo{author}{\bibfnamefont{N.~F.} \bibnamefont{Van~Hulst}},
  \bibinfo{journal}{Light: Science \& Applications}
  \textbf{\bibinfo{volume}{6}}, \bibinfo{pages}{e16239} (\bibinfo{year}{2017}).

\bibitem[{\citenamefont{Brif et~al.}({2010})\citenamefont{Brif, Chakrabarti,
  and Rabitz}}]{brif2010control}
\bibinfo{author}{\bibfnamefont{C.}~\bibnamefont{Brif}},
  \bibinfo{author}{\bibfnamefont{R.}~\bibnamefont{Chakrabarti}},
  \bibnamefont{and} \bibinfo{author}{\bibfnamefont{H.}~\bibnamefont{Rabitz}},
  \bibinfo{journal}{{New J. Phys.}} \textbf{\bibinfo{volume}{{12}}},
  \bibinfo{pages}{{075008}} (\bibinfo{year}{{2010}}).

\bibitem[{\citenamefont{Nuernberger et~al.}({2007})\citenamefont{Nuernberger,
  Vogt, Brixner, and Gerber}}]{nuernberger2007femtosecond}
\bibinfo{author}{\bibfnamefont{P.}~\bibnamefont{Nuernberger}},
  \bibinfo{author}{\bibfnamefont{G.}~\bibnamefont{Vogt}},
  \bibinfo{author}{\bibfnamefont{T.}~\bibnamefont{Brixner}}, \bibnamefont{and}
  \bibinfo{author}{\bibfnamefont{G.}~\bibnamefont{Gerber}},
  \bibinfo{journal}{{Phys. Chem. Chem. Phys.}} \textbf{\bibinfo{volume}{{9}}},
  \bibinfo{pages}{{2470}} (\bibinfo{year}{{2007}}).

\bibitem[{\citenamefont{Shi et~al.}(1988)\citenamefont{Shi, Woody, and
  Rabitz}}]{shi1988optimal}
\bibinfo{author}{\bibfnamefont{S.}~\bibnamefont{Shi}},
  \bibinfo{author}{\bibfnamefont{A.}~\bibnamefont{Woody}}, \bibnamefont{and}
  \bibinfo{author}{\bibfnamefont{H.}~\bibnamefont{Rabitz}},
  \bibinfo{journal}{J. Chem. Phys.} \textbf{\bibinfo{volume}{88}},
  \bibinfo{pages}{6870} (\bibinfo{year}{1988}).

\bibitem[{\citenamefont{Zhu et~al.}(1998)\citenamefont{Zhu, Botina, and
  Rabitz}}]{zhu1998rapidly}
\bibinfo{author}{\bibfnamefont{W.}~\bibnamefont{Zhu}},
  \bibinfo{author}{\bibfnamefont{J.}~\bibnamefont{Botina}}, \bibnamefont{and}
  \bibinfo{author}{\bibfnamefont{H.}~\bibnamefont{Rabitz}},
  \bibinfo{journal}{J. Chem. Phys.} \textbf{\bibinfo{volume}{108}},
  \bibinfo{pages}{1953} (\bibinfo{year}{1998}).

\bibitem[{\citenamefont{Werschnik and Gross}(2007)}]{werschnik2007quantum}
\bibinfo{author}{\bibfnamefont{J.}~\bibnamefont{Werschnik}} \bibnamefont{and}
  \bibinfo{author}{\bibfnamefont{E.}~\bibnamefont{Gross}}, \bibinfo{journal}{J.
  Phys. B: At., Mol. Opt. Phys.} \textbf{\bibinfo{volume}{40}},
  \bibinfo{pages}{R175} (\bibinfo{year}{2007}).

\bibitem[{\citenamefont{Klamroth}(2006)}]{klamroth2006optimal}
\bibinfo{author}{\bibfnamefont{T.}~\bibnamefont{Klamroth}},
  \bibinfo{journal}{J. Chem. Phys.} \textbf{\bibinfo{volume}{124}},
  \bibinfo{pages}{144310} (\bibinfo{year}{2006}).

\bibitem[{\citenamefont{Krause and Klamroth}(2008)}]{krause2008dipole}
\bibinfo{author}{\bibfnamefont{P.}~\bibnamefont{Krause}} \bibnamefont{and}
  \bibinfo{author}{\bibfnamefont{T.}~\bibnamefont{Klamroth}},
  \bibinfo{journal}{J. Chem. Phys.} \textbf{\bibinfo{volume}{128}},
  \bibinfo{pages}{234307} (\bibinfo{year}{2008}).

\bibitem[{\citenamefont{Ohtsuki et~al.}(2003)\citenamefont{Ohtsuki, Nakagami,
  Zhu, and Rabitz}}]{ohtsuki2003quantum}
\bibinfo{author}{\bibfnamefont{Y.}~\bibnamefont{Ohtsuki}},
  \bibinfo{author}{\bibfnamefont{K.}~\bibnamefont{Nakagami}},
  \bibinfo{author}{\bibfnamefont{W.}~\bibnamefont{Zhu}}, \bibnamefont{and}
  \bibinfo{author}{\bibfnamefont{H.}~\bibnamefont{Rabitz}},
  \bibinfo{journal}{Chem. Phys.} \textbf{\bibinfo{volume}{287}},
  \bibinfo{pages}{197} (\bibinfo{year}{2003}).

\bibitem[{\citenamefont{Beyvers et~al.}(2006)\citenamefont{Beyvers, Ohtsuki,
  and Saalfrank}}]{beyvers2006optimal}
\bibinfo{author}{\bibfnamefont{S.}~\bibnamefont{Beyvers}},
  \bibinfo{author}{\bibfnamefont{Y.}~\bibnamefont{Ohtsuki}}, \bibnamefont{and}
  \bibinfo{author}{\bibfnamefont{P.}~\bibnamefont{Saalfrank}},
  \bibinfo{journal}{J. Chem. Phys.} \textbf{\bibinfo{volume}{124}},
  \bibinfo{pages}{234706} (\bibinfo{year}{2006}).

\bibitem[{\citenamefont{Keefer et~al.}(2015)\citenamefont{Keefer, Thallmair,
  Zauleck, and de~Vivie-Riedle}}]{keefer2015multi}
\bibinfo{author}{\bibfnamefont{D.}~\bibnamefont{Keefer}},
  \bibinfo{author}{\bibfnamefont{S.}~\bibnamefont{Thallmair}},
  \bibinfo{author}{\bibfnamefont{J.~P.} \bibnamefont{Zauleck}},
  \bibnamefont{and}
  \bibinfo{author}{\bibfnamefont{R.}~\bibnamefont{de~Vivie-Riedle}},
  \bibinfo{journal}{J. Phys. B: At., Mol. Opt. Phys.}
  \textbf{\bibinfo{volume}{48}}, \bibinfo{pages}{234003}
  (\bibinfo{year}{2015}).

\bibitem[{\citenamefont{Onsager}(1936)}]{Onsager1936}
\bibinfo{author}{\bibfnamefont{L.}~\bibnamefont{Onsager}}, \bibinfo{journal}{J.
  Am. Chem. Soc.} \textbf{\bibinfo{volume}{58}}, \bibinfo{pages}{1486}
  (\bibinfo{year}{1936}).

\bibitem[{\citenamefont{Tomasi et~al.}(2005)\citenamefont{Tomasi, Mennucci, and
  Cammi}}]{tomasi2005quantum}
\bibinfo{author}{\bibfnamefont{J.}~\bibnamefont{Tomasi}},
  \bibinfo{author}{\bibfnamefont{B.}~\bibnamefont{Mennucci}}, \bibnamefont{and}
  \bibinfo{author}{\bibfnamefont{R.}~\bibnamefont{Cammi}},
  \bibinfo{journal}{Chem. Rev.} \textbf{\bibinfo{volume}{105}},
  \bibinfo{pages}{2999} (\bibinfo{year}{2005}).

\bibitem[{\citenamefont{Sklarz and Tannor}(2002)}]{sklarz2002loading}
\bibinfo{author}{\bibfnamefont{S.~E.} \bibnamefont{Sklarz}} \bibnamefont{and}
  \bibinfo{author}{\bibfnamefont{D.~J.} \bibnamefont{Tannor}},
  \bibinfo{journal}{Phys. Rev. A} \textbf{\bibinfo{volume}{66}},
  \bibinfo{pages}{053619} (\bibinfo{year}{2002}).

\bibitem[{\citenamefont{Hohenester et~al.}(2007)\citenamefont{Hohenester,
  Rekdal, Borz{\`\i}, and Schmiedmayer}}]{hohenester2007optimal}
\bibinfo{author}{\bibfnamefont{U.}~\bibnamefont{Hohenester}},
  \bibinfo{author}{\bibfnamefont{P.~K.} \bibnamefont{Rekdal}},
  \bibinfo{author}{\bibfnamefont{A.}~\bibnamefont{Borz{\`\i}}},
  \bibnamefont{and}
  \bibinfo{author}{\bibfnamefont{J.}~\bibnamefont{Schmiedmayer}},
  \bibinfo{journal}{Phys. Rev. A} \textbf{\bibinfo{volume}{75}},
  \bibinfo{pages}{023602} (\bibinfo{year}{2007}).

\bibitem[{\citenamefont{Mundt and Tannor}(2009)}]{mundt2009optimal}
\bibinfo{author}{\bibfnamefont{M.}~\bibnamefont{Mundt}} \bibnamefont{and}
  \bibinfo{author}{\bibfnamefont{D.~J.} \bibnamefont{Tannor}},
  \bibinfo{journal}{New J. Phys.} \textbf{\bibinfo{volume}{11}},
  \bibinfo{pages}{105038} (\bibinfo{year}{2009}).

\bibitem[{\citenamefont{Miertu{\v{s}} et~al.}(1981)\citenamefont{Miertu{\v{s}},
  Scrocco, and Tomasi}}]{Miertus1981}
\bibinfo{author}{\bibfnamefont{S.}~\bibnamefont{Miertu{\v{s}}}},
  \bibinfo{author}{\bibfnamefont{E.}~\bibnamefont{Scrocco}}, \bibnamefont{and}
  \bibinfo{author}{\bibfnamefont{J.}~\bibnamefont{Tomasi}},
  \bibinfo{journal}{Chem. Phys.} \textbf{\bibinfo{volume}{55}},
  \bibinfo{pages}{117} (\bibinfo{year}{1981}).

\bibitem[{\citenamefont{Cammi and Tomasi}(1995{\natexlab{a}})}]{Cammi1995}
\bibinfo{author}{\bibfnamefont{R.}~\bibnamefont{Cammi}} \bibnamefont{and}
  \bibinfo{author}{\bibfnamefont{J.}~\bibnamefont{Tomasi}},
  \bibinfo{journal}{J. Comput. Chem.} \textbf{\bibinfo{volume}{16}},
  \bibinfo{pages}{1449} (\bibinfo{year}{1995}{\natexlab{a}}).

\bibitem[{\citenamefont{Cance`s et~al.}(1997)\citenamefont{Cance`s, Mennucci,
  and Tomasi}}]{Cances1997}
\bibinfo{author}{\bibfnamefont{E.}~\bibnamefont{Cance`s}},
  \bibinfo{author}{\bibfnamefont{B.}~\bibnamefont{Mennucci}}, \bibnamefont{and}
  \bibinfo{author}{\bibfnamefont{J.}~\bibnamefont{Tomasi}},
  \bibinfo{journal}{J. Chem. Phys.} \textbf{\bibinfo{volume}{107}},
  \bibinfo{pages}{3032} (\bibinfo{year}{1997}).

\bibitem[{\citenamefont{Cammi et~al.}(1998)\citenamefont{Cammi, Mennucci, and
  Tomasi}}]{cammi1998calculation}
\bibinfo{author}{\bibfnamefont{R.}~\bibnamefont{Cammi}},
  \bibinfo{author}{\bibfnamefont{B.}~\bibnamefont{Mennucci}}, \bibnamefont{and}
  \bibinfo{author}{\bibfnamefont{J.}~\bibnamefont{Tomasi}},
  \bibinfo{journal}{J. Phys. Chem. A} \textbf{\bibinfo{volume}{102}},
  \bibinfo{pages}{870} (\bibinfo{year}{1998}).

\bibitem[{\citenamefont{Tomasi et~al.}(2002)\citenamefont{Tomasi, Cammi,
  Mennucci, Cappelli, and Corni}}]{tomasi2002molecular}
\bibinfo{author}{\bibfnamefont{J.}~\bibnamefont{Tomasi}},
  \bibinfo{author}{\bibfnamefont{R.}~\bibnamefont{Cammi}},
  \bibinfo{author}{\bibfnamefont{B.}~\bibnamefont{Mennucci}},
  \bibinfo{author}{\bibfnamefont{C.}~\bibnamefont{Cappelli}}, \bibnamefont{and}
  \bibinfo{author}{\bibfnamefont{S.}~\bibnamefont{Corni}},
  \bibinfo{journal}{Phys. Chem. Chem. Phys.} \textbf{\bibinfo{volume}{4}},
  \bibinfo{pages}{5697} (\bibinfo{year}{2002}).

\bibitem[{\citenamefont{Cammi et~al.}(2005)\citenamefont{Cammi, Corni,
  Mennucci, and Tomasi}}]{cammi2005electronic}
\bibinfo{author}{\bibfnamefont{R.}~\bibnamefont{Cammi}},
  \bibinfo{author}{\bibfnamefont{S.}~\bibnamefont{Corni}},
  \bibinfo{author}{\bibfnamefont{B.}~\bibnamefont{Mennucci}}, \bibnamefont{and}
  \bibinfo{author}{\bibfnamefont{J.}~\bibnamefont{Tomasi}},
  \bibinfo{journal}{J. Chem. Phys.} \textbf{\bibinfo{volume}{122}},
  \bibinfo{pages}{104513} (\bibinfo{year}{2005}).

\bibitem[{\citenamefont{Pipolo et~al.}(2014{\natexlab{a}})\citenamefont{Pipolo,
  Corni, and Cammi}}]{Pipolo2014}
\bibinfo{author}{\bibfnamefont{S.}~\bibnamefont{Pipolo}},
  \bibinfo{author}{\bibfnamefont{S.}~\bibnamefont{Corni}}, \bibnamefont{and}
  \bibinfo{author}{\bibfnamefont{R.}~\bibnamefont{Cammi}},
  \bibinfo{journal}{Comput. Theor. Chem.} \textbf{\bibinfo{volume}{1040-1041}},
  \bibinfo{pages}{112} (\bibinfo{year}{2014}{\natexlab{a}}).

\bibitem[{\citenamefont{Tannor}(2007)}]{tannor2007introduction}
\bibinfo{author}{\bibfnamefont{D.}~\bibnamefont{Tannor}},
  \emph{\bibinfo{title}{Introduction to Quantum Mechanics}}
  (\bibinfo{publisher}{University Science Books}, \bibinfo{year}{2007}).

\bibitem[{\citenamefont{Cammi et~al.}(2000)\citenamefont{Cammi, Cappelli,
  Corni, and Tomasi}}]{cammi2000calculation}
\bibinfo{author}{\bibfnamefont{R.}~\bibnamefont{Cammi}},
  \bibinfo{author}{\bibfnamefont{C.}~\bibnamefont{Cappelli}},
  \bibinfo{author}{\bibfnamefont{S.}~\bibnamefont{Corni}}, \bibnamefont{and}
  \bibinfo{author}{\bibfnamefont{J.}~\bibnamefont{Tomasi}},
  \bibinfo{journal}{J. Phys. Chem. A} \textbf{\bibinfo{volume}{104}},
  \bibinfo{pages}{9874} (\bibinfo{year}{2000}).

\bibitem[{\citenamefont{Cammi and Tomasi}(1995{\natexlab{b}})}]{Cammi1995b}
\bibinfo{author}{\bibfnamefont{R.}~\bibnamefont{Cammi}} \bibnamefont{and}
  \bibinfo{author}{\bibfnamefont{J.}~\bibnamefont{Tomasi}},
  \bibinfo{journal}{Int. J. Quantum Chem.} \textbf{\bibinfo{volume}{56}},
  \bibinfo{pages}{465} (\bibinfo{year}{1995}{\natexlab{b}}).

\bibitem[{\citenamefont{Pipolo et~al.}(2017)\citenamefont{Pipolo, Corni, and
  Cammi}}]{pipolo2017equation}
\bibinfo{author}{\bibfnamefont{S.}~\bibnamefont{Pipolo}},
  \bibinfo{author}{\bibfnamefont{S.}~\bibnamefont{Corni}}, \bibnamefont{and}
  \bibinfo{author}{\bibfnamefont{R.}~\bibnamefont{Cammi}}, \bibinfo{journal}{J.
  Chem. Phys.} \textbf{\bibinfo{volume}{146}}, \bibinfo{pages}{064116}
  (\bibinfo{year}{2017}).

\bibitem[{\citenamefont{Corni et~al.}(2014)\citenamefont{Corni, Pipolo, and
  Cammi}}]{Corni2014}
\bibinfo{author}{\bibfnamefont{S.}~\bibnamefont{Corni}},
  \bibinfo{author}{\bibfnamefont{S.}~\bibnamefont{Pipolo}}, \bibnamefont{and}
  \bibinfo{author}{\bibfnamefont{R.}~\bibnamefont{Cammi}}, \bibinfo{journal}{J.
  Phys. Chem. A} \textbf{\bibinfo{volume}{119}}, \bibinfo{pages}{5405}
  (\bibinfo{year}{2014}).

\bibitem[{\citenamefont{Pipolo et~al.}(2014{\natexlab{b}})\citenamefont{Pipolo,
  Corni, and Cammi}}]{pipolo2014cavity}
\bibinfo{author}{\bibfnamefont{S.}~\bibnamefont{Pipolo}},
  \bibinfo{author}{\bibfnamefont{S.}~\bibnamefont{Corni}}, \bibnamefont{and}
  \bibinfo{author}{\bibfnamefont{R.}~\bibnamefont{Cammi}}, \bibinfo{journal}{J.
  Chem. Phys.} \textbf{\bibinfo{volume}{140}}, \bibinfo{pages}{164114}
  (\bibinfo{year}{2014}{\natexlab{b}}).

\bibitem[{\citenamefont{Gil et~al.}(2019)\citenamefont{Gil, Pipolo, Delgado,
  Rozzi, and Corni}}]{gil2019non}
\bibinfo{author}{\bibfnamefont{G.}~\bibnamefont{Gil}},
  \bibinfo{author}{\bibfnamefont{S.}~\bibnamefont{Pipolo}},
  \bibinfo{author}{\bibfnamefont{A.}~\bibnamefont{Delgado}},
  \bibinfo{author}{\bibfnamefont{C.~A.} \bibnamefont{Rozzi}}, \bibnamefont{and}
  \bibinfo{author}{\bibfnamefont{S.}~\bibnamefont{Corni}}, \bibinfo{journal}{J.
  Chem. Theory Comput.} \textbf{\bibinfo{volume}{15}}, \bibinfo{pages}{2306}
  (\bibinfo{year}{2019}).

\bibitem[{\citenamefont{Ohtsuki et~al.}(2004)\citenamefont{Ohtsuki, Turinici,
  and Rabitz}}]{ohtsuki2004generalized}
\bibinfo{author}{\bibfnamefont{Y.}~\bibnamefont{Ohtsuki}},
  \bibinfo{author}{\bibfnamefont{G.}~\bibnamefont{Turinici}}, \bibnamefont{and}
  \bibinfo{author}{\bibfnamefont{H.}~\bibnamefont{Rabitz}},
  \bibinfo{journal}{J. Chem. Phys.} \textbf{\bibinfo{volume}{120}},
  \bibinfo{pages}{5509} (\bibinfo{year}{2004}).

\bibitem[{\citenamefont{{P{\'{e}}rez Lustres}
  et~al.}(2005)\citenamefont{{P{\'{e}}rez Lustres}, Kovalenko, Mosquera,
  Senyushkina, Flasche, and Ernsting}}]{PerezLustres2005}
\bibinfo{author}{\bibfnamefont{J.~L.} \bibnamefont{{P{\'{e}}rez Lustres}}},
  \bibinfo{author}{\bibfnamefont{S.~A.} \bibnamefont{Kovalenko}},
  \bibinfo{author}{\bibfnamefont{M.}~\bibnamefont{Mosquera}},
  \bibinfo{author}{\bibfnamefont{T.}~\bibnamefont{Senyushkina}},
  \bibinfo{author}{\bibfnamefont{W.}~\bibnamefont{Flasche}}, \bibnamefont{and}
  \bibinfo{author}{\bibfnamefont{N.~P.} \bibnamefont{Ernsting}},
  \bibinfo{journal}{Angew. Chem. Int. Ed.} \textbf{\bibinfo{volume}{44}},
  \bibinfo{pages}{5635} (\bibinfo{year}{2005}).

\bibitem[{\citenamefont{Frisch et~al.}()\citenamefont{Frisch, Trucks, Schlegel,
  Scuseria, Robb, Cheeseman, Scalmani, Barone, Mennucci, Petersson
  et~al.}}]{frisch2009gaussian}
\bibinfo{author}{\bibfnamefont{M.~J.} \bibnamefont{Frisch}},
  \bibinfo{author}{\bibfnamefont{G.~W.} \bibnamefont{Trucks}},
  \bibinfo{author}{\bibfnamefont{H.~B.} \bibnamefont{Schlegel}},
  \bibinfo{author}{\bibfnamefont{G.~E.} \bibnamefont{Scuseria}},
  \bibinfo{author}{\bibfnamefont{M.~A.} \bibnamefont{Robb}},
  \bibinfo{author}{\bibfnamefont{J.~R.} \bibnamefont{Cheeseman}},
  \bibinfo{author}{\bibfnamefont{G.}~\bibnamefont{Scalmani}},
  \bibinfo{author}{\bibfnamefont{V.}~\bibnamefont{Barone}},
  \bibinfo{author}{\bibfnamefont{B.}~\bibnamefont{Mennucci}},
  \bibinfo{author}{\bibfnamefont{G.~A.} \bibnamefont{Petersson}},
  \bibnamefont{et~al.}, \emph{\bibinfo{title}{Gaussian~09}},
  \bibinfo{note}{gaussian Inc. Wallingford CT 2009}.

\bibitem[{\citenamefont{Schmidt et~al.}(1993)\citenamefont{Schmidt, Baldridge,
  Boatz, Elbert, Gordon, Jensen, Koseki, Matsunaga, Nguyen, Su
  et~al.}}]{schmidt1993general}
\bibinfo{author}{\bibfnamefont{M.~W.} \bibnamefont{Schmidt}},
  \bibinfo{author}{\bibfnamefont{K.~K.} \bibnamefont{Baldridge}},
  \bibinfo{author}{\bibfnamefont{J.~A.} \bibnamefont{Boatz}},
  \bibinfo{author}{\bibfnamefont{S.~T.} \bibnamefont{Elbert}},
  \bibinfo{author}{\bibfnamefont{M.~S.} \bibnamefont{Gordon}},
  \bibinfo{author}{\bibfnamefont{J.~H.} \bibnamefont{Jensen}},
  \bibinfo{author}{\bibfnamefont{S.}~\bibnamefont{Koseki}},
  \bibinfo{author}{\bibfnamefont{N.}~\bibnamefont{Matsunaga}},
  \bibinfo{author}{\bibfnamefont{K.~A.} \bibnamefont{Nguyen}},
  \bibinfo{author}{\bibfnamefont{S.}~\bibnamefont{Su}}, \bibnamefont{et~al.},
  \bibinfo{journal}{J. Comput. Chem.} \textbf{\bibinfo{volume}{14}},
  \bibinfo{pages}{1347} (\bibinfo{year}{1993}).

\end{thebibliography}

\end{document}